\documentclass{article}
\usepackage{amsfonts}
\usepackage{amsmath}
\usepackage{amssymb}
\usepackage{epsfig}
\usepackage{float}

\def\beq{\begin{equation}}
\def\eeq{\end{equation}}
\def\bq{\begin{quote}}
\def\eq{\end{quote}}

\def\simlt{\stackrel{<}{{}_\sim}}
\def\simgt{\stackrel{>}{{}_\sim}}

\newcommand{\newc}{\newcommand*}

\long\def\begincomment#1\endcomment{%
        \begingroup\sf\baselineskip12pt#1\endgroup}

\newc{\etal}{\textrm{et al.}} 
\newc{\eg}{\textrm{e.g.}} 
\newc{\ie}{\textrm{i.e.}}
\newc{\etc}{\textrm{etc}}
\newc\vs{\textrm{vs.}}
\newc{\cl}{\rm {C.L.}}
\newc{\ev}{\ensuremath{\,\mathrm{eV}}}
\newc{\kev}{\ensuremath{\,\mathrm{keV}}}
\newc{\mev}{\ensuremath{\,\mathrm{MeV}}}
\newc{\gev}{\ensuremath{\,\mathrm{GeV}}}
\newc{\tev}{\ensuremath{\,\mathrm{TeV}}}
\newc{\MeV}{\mev} 
\newc{\TeV}{\tev}
\newc{\invpb}{\ensuremath{/\text{pb}}}
\newc{\invfb}{\ensuremath{/\text{fb}}}
\newc\nb{\ensuremath{\,\mathrm{nb}}} \newc\pb{\ensuremath{\,\mathrm{pb}}} \newc\fb{\ensuremath{\,\mathrm{fb}}}
\newc\pc{\ensuremath{\,\mathrm{pc}}}
\newc\kpc{\ensuremath{\,\mathrm{kpc}}}
\newc\mpc{\ensuremath{\,\mathrm{Mpc}}}
\newc\ps{\ensuremath{\,\mathrm{ps}}} 
\newc\cmeter{\ensuremath{\,\mathrm{cm}}} 
\newc\meter{\ensuremath{\,\mathrm{m}}} 
\newc\kmeter{\ensuremath{\,\mathrm{km}}}
\newc\second{\ensuremath{\,\mathrm{s}}}
\newc\msecond{\ensuremath{\,\mathrm{ms}}}
\newc\nsecond{\ensuremath{\,\mathrm{ns}}}
\newc\psecond{\ensuremath{\,\mathrm{ps}}}
\newc{\chisqmin}{\ensuremath{\chi^2_{\mathrm{min}}}}
\newc{\Delchisq}{\ensuremath{\Delta\chi^2}}
\newc{\chisq}{\ensuremath{\chi^2}}
\newc{\like}{\ensuremath{\mathcal{L}}}
\newc\lsim{\ensuremath{\mathrel{\rlap{\lower4pt\hbox{\hskip1pt$\sim$}}\raise1pt\hbox{$<$}}}}
\newc\gsim{\ensuremath{\mathrel{\rlap{\lower4pt\hbox{\hskip1pt$\sim$}}\raise1pt\hbox{$>$}}}}
\newc{\VEV}[1]{\ensuremath{\langle #1 \rangle}}
\newc{\dl}{\ensuremath{\stackrel{\leftarrow}{D}}}
\newc{\dr}{\ensuremath{\stackrel{\rightarrow}{D}}}
\newc{\bcenter}{\begin{center}}    \newc{\ecenter}{\end{center}}
\newc{\bfl}{\begin{flushleft}}    \newc{\efl}{\end{flushleft}}
\newc{\bfr}{\begin{flushright}}    \newc{\efr}{\end{flushright}}

\newc{\bi}{\begin{itemize}}
\newc{\ei}{\end{itemize}}
\newc{\bed}{\begin{description}}
\newc{\eed}{\end{description}}
\newc{\ben}{\begin{enumerate}}
\newc{\een}{\end{enumerate}}

\newc{\be}{\begin{equation}}
\newc{\ee}{\end{equation}}
\newc{\bea}{\begin{eqnarray}}
\newc{\eea}{\end{eqnarray}}
\newc{\ra}{\rightarrow}
\newc{\alphas}{\ensuremath{\alpha_s}}
\newc{\alphatwo}{\ensuremath{\alpha_2}}
\newc{\alphaone}{\ensuremath{\alpha_1}}
\newc{\alphai}[1]{\ensuremath{\alpha_{#1}}}
\newc{\alphaem}{\ensuremath{\alpha_{\mathrm{em}}}}
\newc{\alphaeff}{\ensuremath{\alpha_{\mathrm{eff}}}}
\newc{\sineff}{\ensuremath{\sin \theta_{\mathrm{eff}}}}
\newc{\sinsqeff}{\ensuremath{\sin^2 \theta_{\mathrm{eff}}}}
\newc{\dalphahad}{\ensuremath{\Delta \alpha_{\mathrm{had}}}}
\newc{\yt}{\ensuremath{h_t}} \newc{\yb}{\ensuremath{h_b}} \newc{\ytau}{\ensuremath{h_{\tau}}}
\newc\mz{\ensuremath{M_Z}} 
\newc\mw{\ensuremath{m_W}}
\newc\mZ{\mz}        \newc\mW{\mw}
\newc\mhsm{\ensuremath{ m_{H_{\mathrm{SM}}}}}
\newc{\mtop}{\ensuremath{ m_t}}               \newc{\mtpole}{\ensuremath{ M_t}}
\newc{\mbottom}{\ensuremath{ m_b}} 
\newc{\mtau}{\ensuremath{ m_{\tau}}}
\newc{\mt}{\mtpole}
\newc{\mb}{\mbottom} 
\newc{\rgg}{\ensuremath{R_{h}(\gamma\gamma)}}
\newc{\rzz}{\ensuremath{R_{h}(ZZ)}}
\newc{\rtwogg}{\ensuremath{R_{h_2}(\gamma\gamma)}}
\newc{\rtwozz}{\ensuremath{R_{h_2}(ZZ)}}
\newc{\ronegg}{\ensuremath{R_{h_1}(\gamma\gamma)}}
\newc{\ronezz}{\ensuremath{R_{h_1}(ZZ)}}
\newc{\rsiggg}{\ensuremath{R_{h_\textrm{sig}}(\gamma\gamma)}}
\newc{\rsigzz}{\ensuremath{R_{h_\textrm{sig}}(ZZ)}}
\newc{\llbar}{\ensuremath{\ell\bar{\ell}}}
\newc{\tauptaum}{\ensuremath{ \tau^+\tau^-}}
\newc{\qqbar}{\ensuremath{ q\bar{q}}} \newc{\ppbar}{\ensuremath{ p\bar{p}}}
\newc{\bbbar}{\ensuremath{ b\bar{b}}} \newc{\ttbar}{\ensuremath{ t\bar{t}}}
\newc{\ffbar}{\ensuremath{ f\bar{f}}} \newc{\tautaubar}{\ensuremath{ \tau\bar{\tau}}}
\newc{\mchi}{\ensuremath{m_{\chi}}}
\newc{\squark}{\ensuremath{\tilde{q}}}
\newc{\slepton}{\ensuremath{\tilde{l}}}
\newc{\gluino}{\ensuremath{\tilde{g}}} 
\newc{\mgluino}{\ensuremath{{m_{\gluino}}}}
\newc{\tone}{\ensuremath{{\tilde{t}_1}}}
\newc{\sthw}{\ensuremath{ \sin\theta_W}}              \newc{\cthw}{\ensuremath{\cos\theta_W}}
\newc{\tanthw}{\ensuremath{ \tan\theta_W}}              \newc{\cotthw}{\ensuremath{\cot\theta_W}}
\newc{\ssqthw}{\ensuremath{\sin^2 \theta_W}}
\newc{\msbar}{\ensuremath{\overline{MS}}} \newc{\drbar}{\ensuremath{\overline{DR}}}
\newc{\mtmtsmmsbar}{\ensuremath{ m_t(m_t)^{\msbar}_{{\mathrm{SM}}}}}
\newc{\mtmtsmdrbar}{\ensuremath{ m_t(m_t)^{\drbar}_{{\mathrm{SM}}}}}
\newc{\mtmtmssmdrbar}{\ensuremath{ m_t(m_t)^{\drbar}_{{\mathrm{SUSY}}}}}
\newc{\mbmbmsbar}{\ensuremath{ m_b(m_b)^{\msbar} }}
\newc{\mbmbsmmsbar}{\ensuremath{ m_b(m_b)^{\msbar}_{{\mathrm{SM}}}}}
\newc{\mbmzsmmsbar}{\ensuremath{ m_b(\mz)^{\msbar}_{{\mathrm{SM}}}}}
\newc{\mbmzsmdrbar}{\ensuremath{ m_b(\mz)^{\drbar}_{{\mathrm{SM}}}}}
\newc{\mbmzmssmdrbar}{\ensuremath{ m_b(\mz)^{\drbar}_{{\mathrm{SUSY}}}}}
\newc{\mtaumzsmmsbar}{\ensuremath{ m_{\tau}(\mz)^{\msbar}_{{\mathrm{SM}}}}}
\newc{\mtaumzsmdrbar}{\ensuremath{ m_{\tau}(\mz)^{\drbar}_{{\mathrm{SM}}}}}
\newc{\mtaumzmssmdrbar}{\ensuremath{ m_{\tau}(\mz)^{\drbar}_{{\mathrm{SUSY}}}}}
\newc{\alphasmzms}{\ensuremath{\alpha_s(M_Z)^{\overline{MS}}}}
\newc{\alphaimzms}[1]{\ensuremath{\alpha_{#1}(M_Z)^{\overline{MS}}}}
\newc{\alphaemmz}{\ensuremath{\alpha_{\mathrm{em}}(M_Z)^{\overline{MS}}}}
\newc{\mzero}{\ensuremath{{m_0}}}
\newc{\mhalf}{\ensuremath{ m_{1/2}}}
\newc{\tanb}{\ensuremath{\tan\beta}}
\newc{\azero}{\ensuremath{ A_0}}
\newc{\bzero}{\ensuremath{ B_0}}
\newc{\signmu}{\ensuremath{\rm{sgn}\,\mu}}
\newc{\mueff}{\ensuremath{\mu_{\rm{eff}}}}
\newc{\lam}{\ensuremath{{\lambda}}}
\newc{\kap}{\ensuremath{{\kappa}}}
\newc{\alam}{\ensuremath{{A_{\lambda}}}}
\newc{\akap}{\ensuremath{{A_{\kappa}}}}
\newc{\hs}{\ensuremath{ H_s}}      
\newc{\mhs}{\ensuremath{ m_{H_s}}} 
\newc{\mgut}{\ensuremath{ M_{\rm GUT}}}
\newc{\mplanck}{\ensuremath{ M_{\rm P}}}      \newc{\mpl}{\ensuremath{ M_{\rm Pl}}}
\newc{\msusy}{\ensuremath{ M_{\rm SUSY}}}      \newc{\ms}{\ensuremath{ M_{\rm S}}}
 \newc{\mhl}{\ensuremath{m_\hl}} 
 \newc{\mhone}{\ensuremath{m_{h_1}}} 
 \newc{\mhtwo}{\ensuremath{m_{h_2}}} 
 \newc{\mglu}{\ensuremath{m_{\tilde g}}} 
 \newc{\mul}{\ensuremath{m_{\tilde{u}_L}}} 
 \newc{\mtone}{\ensuremath{m_{\tilde{t}_1}}} 
 \newc{\ma}{\ensuremath{m_A}} 
 \newc{\maone}{\ensuremath{m_{a_1}}} 
 \newc{\matwo}{\ensuremath{m_{a_2}}}
 \newc{\hone}{\ensuremath{h_1}}
 \newc{\htwo}{\ensuremath{h_2}}
 \newc{\aone}{\ensuremath{a_1}}
 \newc{\atwo}{\ensuremath{a_2}}
 \newc{\mhu}{\ensuremath{ m_{H_u}}}       
 \newc{\mhd}{\ensuremath{ m_{H_d}}}
 \newc{\mhusq}{\ensuremath{ m_{H_u}^2}}       
 \newc{\mhdsq}{\ensuremath{ m_{H_d}^2}}
 \newc{\mhuew}{\ensuremath{ m^{\ast}_{H_u}}}       
 \newc{\mhdew}{\ensuremath{ m^{\ast}_{H_d}}}
 \newc{\mhuewsq}{\ensuremath{ m^{\ast\, 2}_{H_u}}}       
 \newc{\mhdewsq}{\ensuremath{ m^{\ast\, 2}_{H_d}}}
 \newc{\hu}{\ensuremath{ H_u}}       
 \newc{\hd}{\ensuremath{ H_d}}

 \newc{\mqthree}{\ensuremath{m_{\widetilde{Q}_3}^2}}
 \newc{\muthree}{\ensuremath{m_{\tilde{u}_3}^2}}
 \newc{\mdthree}{\ensuremath{m_{\tilde{d}_3}^2}}
 \newc{\mlthree}{\ensuremath{m_{\widetilde{L}_3}^2}}
 \newc{\methree}{\ensuremath{m_{\tilde{e}_3}^2}}
 \newc{\mqtwo}{\ensuremath{m_{\widetilde{Q}_2}^2}}
 \newc{\mutwo}{\ensuremath{m_{\tilde{u}_2}^2}}
 \newc{\mdtwo}{\ensuremath{m_{\tilde{d}_2}^2}}
 \newc{\mltwo}{\ensuremath{m_{\widetilde{L}_2}^2}}
 \newc{\metwo}{\ensuremath{m_{\tilde{e}_2}^2}}
 \newc{\mqone}{\ensuremath{m_{\widetilde{Q}_1}^2}}
 \newc{\muone}{\ensuremath{m_{\tilde{u}_1}^2}}
 \newc{\mdone}{\ensuremath{m_{\tilde{d}_1}^2}}
 \newc{\mlone}{\ensuremath{m_{\widetilde{L}_1}^2}}
 \newc{\meone}{\ensuremath{m_{\tilde{e}_1}^2}}
 \newc{\mone}{\ensuremath{M_1}}
 \newc{\monesq}{\ensuremath{M_1^2}}
 \newc{\mtwo}{\ensuremath{M_2}}
 \newc{\mtwosq}{\ensuremath{M_2^2}}
 \newc{\mthree}{\ensuremath{M_3}}
 \newc{\mthreesq}{\ensuremath{M_3^2}}
 \newc{\atau}{\ensuremath{{A_{\tau}}}}
 \newc{\at}{\ensuremath{{A_{t}}}}
 \newc{\ab}{\ensuremath{{A_{b}}}}
 \newc{\atausq}{\ensuremath{{A_{\tau}^2}}}
 \newc{\atsq}{\ensuremath{{A_{t}^2}}}
 \newc{\absq}{\ensuremath{{A_{b}^2}}}
 \newc{\dmzero}{\ensuremath{\Delta{_{m_0}}}}
 \newc{\dmhalf}{\ensuremath{\Delta{_{m_{1/2}}}}}
 \newc{\dmu}{\ensuremath{\Delta{_{\mu}}}}

 \newc{\pten}{\ensuremath{\psi_{10}}}
 \newc{\ffive}{\ensuremath{\phi_{5}}}
 \newc{\hfive}{\ensuremath{h_{5}}}
 \newc{\hbfive}{\ensuremath{h_{\bar{5}}}}
 \newc{\thet}{\ensuremath{\theta_{50}}}
 \newc{\thetb}{\ensuremath{\theta_{\,\overline{50}}}}
 \newc{\ptenhat}{\ensuremath{\hat{\psi}_{10}}}
 \newc{\ffivehat}{\ensuremath{\hat{\phi}_{5}}}
 \newc{\hfivehat}{\ensuremath{\hat{h}_{5}}}
 \newc{\hbfivehat}{\ensuremath{\hat{h}_{\bar{5}}}}
 \newc{\thethat}{\ensuremath{\hat{\theta}_{50}}}
 \newc{\thetbhat}{\ensuremath{\hat{\theta}_{\,\overline{50}}}}
 \newc{\si}{\ensuremath{\Sigma}}
 \newc{\mfive}{\ensuremath{m_5^2}}
 \newc{\mten}{\ensuremath{m_{10}^2}}
 \newc{\dfive}{\ensuremath{\Delta^2_5}}
 \newc{\dbfive}{\ensuremath{\Delta^2_{\bar{5}}}}
 \newc{\dfifty}{\ensuremath{\Delta^2_{50}}}
 \newc{\dfiftyb}{\ensuremath{\Delta^2_{\,\overline{50}}}}
 \newc{\msi}{\ensuremath{m_{\Sigma}^2}}
 \newc{\lamh}{\ensuremath{\lambda_{H}}}
 \newc{\lamhb}{\ensuremath{\lambda_{\bar{H}}}}
 \newc{\ah}{\ensuremath{A_{H}}}
 \newc{\ahb}{\ensuremath{A_{\bar{H}}}}
 \newc{\lams}{\ensuremath{\lambda_{S}}}
 \newc{\as}{\ensuremath{A_{S}}}
 \newc{\lamsig}{\ensuremath{\lambda_{\si}}}
 \newc{\asig}{\ensuremath{A_{\si}}}

 \newc{\msten}{\ensuremath{m_{16}^2}}
 \newc{\mhun}{\ensuremath{m_{126}^2}}
 \newc{\mhunb}{\ensuremath{m_{\bar{126}}^2}}
 \newc{\mthun}{\ensuremath{m_{210}^2}}
 \newc{\ahun}{\ensuremath{A_{\bar{126}}}}
 \newc{\yhun}{\ensuremath{Y_{\bar{126}}}}
 \newc{\aten}{\ensuremath{A_{10}}}
 \newc{\yten}{\ensuremath{Y_{10}}}
 \newc{\alone}{\ensuremath{A_{\lambda_1}}}
 \newc{\altwo}{\ensuremath{A_{\lambda_2}}}
 \newc{\althree}{\ensuremath{A_{\lambda_3}}}
 \newc{\althreeb}{\ensuremath{A_{\bar{\lambda_3}}}}
 \newc{\lone}{\ensuremath{\lambda_1}}
 \newc{\ltwo}{\ensuremath{\lambda_2}}
 \newc{\lthree}{\ensuremath{\lambda_3}}
 \newc{\lthreeb}{\ensuremath{\bar{\lambda_3}}}

\newc{\sigsip}{\ensuremath{\sigma^{\rm SI}_{p}}}	\newc{\sigsin}{\ensuremath{\sigma^{\rm SI}_{n}}}
\newc{\sigsdp}{\ensuremath{\sigma^{\rm SD}_{p}}}	\newc{\sigsdn}{\ensuremath{\sigma^{\rm SD}_{n}}}
\newc{\sigsi}{\ensuremath{\sigma^{\rm SI}}}	\newc{\sigsd}{\ensuremath{\sigma^{\rm SD}}}
\newc{\abund}{\ensuremath{ \Omega h^2}}
\newc{\omegadm}{\ensuremath{ \Omega_{{\rm DM}}}}     \newc{\abunddm}{\ensuremath{ \Omega_{{\rm DM}} h^2}} 
\newc{\omegam}{\ensuremath{ \Omega_{{\rm m}}}}       \newc{\abundm}{\ensuremath{ \Omega_{{\rm m}} h^2}}
\newc{\omegab}{\ensuremath{ \Omega_{{\rm b}}}}	\newc{\abundb}{\ensuremath{ \Omega_{{\rm b}} h^2}}
\newc{\omegatot}{\ensuremath{ \Omega_{{\rm TOT}}}}
\newc{\omegacdm}{\ensuremath{ \Omega_{{\rm CDM}}}}   \newc{\abundcdm}{\ensuremath{ \Omega_{{\rm CDM}} h^2}}
\newc{\omegalambda}{\ensuremath{ \Omega_{\Lambda}}} \newc{\abundlambda}{\ensuremath{ \Omega_{\Lambda} h^2}}
\newc{\omegarad}{\ensuremath{ \Omega_{{\rm rad}}}}  \newc{\abundrad}{\ensuremath{ \Omega_{{\rm rad}} h^2}}
\newc{\rhocrit}{\ensuremath{ \rho_{\rm crit}}}
\newc{\rhochi}{\ensuremath{ \rho_{\chi}}}
\newc{\abunchi}{\ensuremath{\Omega_\chi h^2}}
\newc{\abundlsp}{\ensuremath{\Omega_{\rm LSP}h^2}}
\newc{\amu}{\ensuremath{ a_{\mu}}}        \newc{\amususy}{\ensuremath{ a_{\mu}^{\mathrm{SUSY}}}}
\newc{\amuexpt}{\ensuremath{ a_{\mu}^{\mathrm{expt}}}}        \newc{\amusm}{\ensuremath{ a_{\mu}^{\mathrm{SM}}}}
\newc\deltaamu{\ensuremath{\Delta a_{\mu}}} \newc{\deltaamususy}{\ensuremath{\delta a_{\mu}^{\mathrm{SUSY}}}}
\newc\gmtwo{\ensuremath{ (g-2)_{\mu}}} 
\newc{\deltagmtwomususy}{\ensuremath{\delta\left(g-2\right)_{\mu}^{\mathrm{SUSY}}}}
\newc{\deltagmtwomu}{\ensuremath{\delta\left(g-2\right)_{\mu}}}
\newc\BR{\ensuremath{\rm BR}}
\newc\bsgamma{\ensuremath{ b\rightarrow s \gamma }}
\newc\bxsgamma{\ensuremath{\overline{B}\rightarrow X_{s}\gamma}}
\newc\brbsgamma{\ensuremath{\BR\left(\bsgamma\right)}}
\newc\brbxsgamma{\ensuremath{\BR\left(\bxsgamma\right)}}
\newc\bsmumu{\ensuremath{B_s\to\mu^+\mu^-}}
\newc\brbsmumu{\ensuremath{\BR\left(B_s\to\mu^+\mu^-\right)}}
\newc\bdmmumu{\ensuremath{\overline{B}_d\to\mu^+\mu^-}}
\newc\bbbarmix{\ensuremath{\overline{B}_s\mbox{-}B_s}}      % B_s mixing
\newc\delmbs{\ensuremath{\Delta M_{B_s}}}
\newc{\butaunu}{\ensuremath{B_u \rightarrow \tau \nu}}
\newc{\brbutaunu}{\ensuremath{\BR\left(B_u \rightarrow \tau \nu\right)}}
\let\oldcite\cite
\renewcommand*{\cite}{~\oldcite}

\newcommand*{\hl}{\ensuremath{h}}

\newcommand*{\madgr}{\texttt{MadGraph5\_aMC@NLO}}
\begin{document}

\begin{titlepage}

\noindent
\begin{flushright}
%lr *** preprint number \\
\end{flushright}
\vspace{1cm}

\begin{center}
  \begin{Large}
    \begin{bf}
Axino dark matter with low reheating temperature
        \end{bf}
  \end{Large}
\end{center}

\vspace{0.5cm}
\begin{center}
\begin{large}
Leszek Roszkowski,$^{a}$\footnote{On leave of absence from the
  University of Sheffield, U.K.}
 Sebastian Trojanowski,$^{a}$ Krzysztof Turzy\'nski$^{b}$
\end{large}

\vspace{0.3cm}
\begin{it}
${}^{a}$ National Centre for Nuclear Research, Ho\.za 69, 00-681, Warsaw, Poland \\
\vspace{0.1cm}
${}^{b}$Institute of Theoretical Physics, Faculty of Physics, University of Warsaw, \\Pasteura 5, 02-093, Warsaw, Poland\\
\end{it}

\vspace{1cm}
\end{center}

\begin{abstract}
We examine  axino dark matter in the regime of a low reheating temperature, $T_R$, after inflation and taking into account that reheating is a non-instantaneous process. This can have a significant effect on the dark matter abundance, mainly due to entropy production in inflaton decays. We study both thermal and non-thermal production of axinos in the framework of the MSSM with ten free parameters. We identify the ranges of the axino mass and the reheating temperature allowed by the LHC and other particle physics data in different models of axino interactions. We confront these limits with cosmological constraints coming the observed dark matter density, large structures formation and big bang nucleosynthesis. We find a number of differences in the phenomenologically acceptable values of the axino mass $m_{\tilde a}$ and the reheating temperature relative to previous studies. In particular, an upper bound on $m_{\tilde a}$ becomes dependent on $T_R$, reaching a maximum value at $T_R\simeq 10^2\,\mathrm{GeV}$. If the lightest ordinary supersymmetric particle is a wino or a higgsino, we obtain a lower limit of approximately $10\,\mathrm{GeV}$ for the reheating temperature. We demonstrate also that entropy production during reheating affects the maximum allowed axino mass and lowest values of the reheating temperature.

\end{abstract}

\vspace{3cm}

\end{titlepage}

%%%%%%%%%%%%%%%%%%%%%%%%%

\tableofcontents

\section{Introduction}
\label{sec:intro}

A variety of cosmological data points to the existence of a non-luminous component of the matter in the Universe,
referred to as dark matter (DM). In spite of decades-long efforts aiming at detecting DM particles directly, they have so far remained elusive.
Indirect probes, such as the impact of DM on the formation and evolution of large structures in the Universe or on the spectrum of the temperature fluctuations in the cosmic microwave background provide powerful tools to study DM. However, such probes involve only some properties of DM particles, 
such as
their (time-dependent) energy density and free-streaming length, hence the masses of proposed DM candidates span thirty orders of magnitude while their cross-sections for scattering with known particles -- forty orders of magnitude (see, e.g.,~\cite{Baer:2014eja} for a recent review). 

There are many examples of DM candidates which were not introduced {\em ad hoc}, but possess a sound theoretical motivation. They include bosonic fields in coherent motion, such as axions, weakly interacting massive particles (WIMPs), such as the lightest neutralino of the Minimal Supersymmetric Standard Model (MSSM), or extremely weakly interacting particles (EWIMPs), such as gravitinos and axinos.

The axion was originally introduced as a solution to the so-called strong CP problem. The smallness of the electric dipole moment of
the neutron can be understood if there is a $U(1)_\mathrm{PQ}$ symmetry, referred to as Peccei-Quinn (PQ) symmetry \cite{Peccei:1977hh,Peccei:1977ur}, which is spontaneously broken
at a scale $f_a$. The very light pseudo-Goldstone boson $a$ associated with this symmetry, called an axion,
couples to the gluon anomaly \cite{Weinberg:1977ma}  (see also, e.g.,~\cite{Kim:2008hd} for a review),
\beq
\label{eq:axion}
\mathcal{L}_\mathrm{axion} = \frac{\alpha_s}{8\pi f_a} a\,G^a_{\mu\nu}\widetilde{G}^{a\,\mu\nu}\,,
\eeq
where $\alpha_s=g_s^2/4\pi$ and $g_s$ is the strong coupling constant.
Various cosmological considerations suggest that $f_a$ lies between approximately $10^{9}\,\mathrm{GeV}$ and $10^{12}\,\mathrm{GeV}$
\cite{Bae:2008ue}. 

Two broad frameworks for field theory models of axions have been considered in the literature. In the Kim-Shifman-Vainshtein-Zakharov (KSVZ) model \cite{Kim:1979if,Shifman:1979if}, the gluon anomaly term (\ref{eq:axion}) is induced through a loop of very heavy singlet quarks carrying PQ charges, while in the Dine-Fischler-Srednicki-Zhitnitsky (DFSZ) model \cite{Dine:1981rt,Zhitnitsky:1980tq} the PQ charges are assigned to the Standard Model fields. These assumptions are sufficient to determine unambiguously axion interactions with the Standard Model fields that are relevant for cosmology, 
though specific implementations of these models can be rather complicated \cite{Kim:2008hd,Kim:1998va}.  
The Lagrangian describing axion interactions with Standard Model particles is given in Appendix A.

In supersymmetric models \cite{Martin:1997} the axion resides in a chiral multiplet with a fermionic superpartner -- the axino $\tilde{a}$ -- whose interactions with the Standard Model particles are related to those of the axion through supersymmetry. See \cite{Baer:2014eja,Choi:2013} for a review, relevant formulae are
shown in Appendix A. The properties of the axino in models with softly broken supersymmetry can be relevant for cosmology \cite{Nilles:1981py,Tamvakis:1982mw,Frere:1982sg}, especially when it is the lightest supersymmetric particle (LSP) and constitutes cold DM \cite{Covi:1999ty,Covi:2001nw}. Particular scenarios of supersymmetry breaking give predictions for the axino mass \cite{Chun:1992zk,Chun:1995hc,Kim:2012bb}; however, due to a strong model dependence and  the absence of an universally accepted scheme of supersymmetry breaking, 
we adopt a phenomenological approach and treat the axino mass as a free parameter.

The mechanisms for axino generation in the post-inflationary Universe (see \cite{Covi:2009pq} for a review) are analogous to those for the gravitino: thermal production (TP) from scatterings and decays of other particles in thermal equilibrium and non-thermal production (NTP) from out-of-equilibrium decays of heavier particles. Although detailed predictions for the axino abundance from TP, $Y_{\tilde{a}}^\mathrm{TP}$, strongly depend on the model of axion interactions, it is inversely proportional to $f_a^2$ and does not depend on the axino mass. In this respect, the axino
differs significantly from the gravitino: as the latter contains a goldstino component related to broken supersymmetry,  the gravitino abundance from TP is inversely proportional to $M_P^2$ and to the square of the gravitino mass. 
A notable difference between the axino interaction models is that in KSVZ models $Y_{\tilde{a}}^\mathrm{TP}$ is proportional to  the reheating temperature $T_R$ defined in terms of the inflaton decay rate $\Gamma_{\phi} = \pi\sqrt{g_{\ast}(T_R)/90}\,(T_R^2/M_{\textrm{Pl}})$, while in DFSZ models $Y_{\tilde{a}}^\mathrm{TP} $ does not depend on $T_R$.
This can be attributed to the fact that in DFSZ models axinos are mainly
produced from decays of thermal particles, while in KSVZ models the main source of axinos are scatterings of strongly interacting particles.
Existing numerical analyses \cite{Brandenburg:2004du,Chun:2011zd} suggest that, in order not to overclose the Universe in the KSVZ scheme, $T_R$ should be at most a few orders of magnitude larger than the electroweak scale.

The contribution of the NTP  to the axino abundance comes from decays of the lightest ordinary supersymmetric particles (LOSP) which underwent freeze-out,
hence $Y_{\tilde{a}}^\mathrm{NTP}=Y_{\mathrm{LOSP}}$, or in terms of cosmological parameters, 
\beq
\Omega_{\tilde{a}}^\mathrm{NTP}h^2 = \frac{m_{\tilde{a}}}{m_\mathrm{LOSP}}\,\Omega_\mathrm{LOSP}h^2.
\label{eq:Oh2ntp}
\eeq 
Depending on the masses of the axino and the LOSP, the LOSP lifetime may be so long (above seconds) that
the highly energetic LOSP decay products can affect successful predictions of the big bang nucleosynthesis (BBN). 
 
Although there has been remarkable progress in the understanding of axino physics and cosmology in the last two decades, the latest LHC data and the increasing precision of reconstructing the history of the Universe motivate extending the existing analyses in two ways. Previous studies of axino DM focused on axino interactions, while some specific assumption about the MSSM mass spectra were made
for simplicity. 
Also, the low $T_R$ regime was studied assuming instantaneous reheating and some fixed typical values of the abundance of
axino DM originating from NTP.
Given many constraints that the LHC data put on the MSSM, it is worthwhile to 
ask what are the allowed and excluded ranges of the parameters of the general MSSM with axino DM.
It is also known that the altered expansion rate of the Universe during reheating may result in a significant change of DM abundance \cite{Giudice:2000,Fornengo:2002,Gelmini:2006Feb,Gelmini:2006May,Gelmini:2006Oct,Strumia:2010aa,Roszkowski:2014lga,Co:2015pka}. 

In this paper, we address the two issues mentioned above.
 We identify the phenomenologically viable
parameter ranges of the MSSM with axino DM and 
calculate accurately the axino abundance \cite{Choi:2011yf} 
not only during the radiation dominated (RD) period, but also during the phase of reheating after inflation \cite{Roszkowski:2014lga}. 
The paper is organized as follows. In Section \ref{sec:axi1}, we 
examine how non-instantaneous reheating affects the predictions for axino DM. In Section \ref{sec:axi2}, we
show the results of our 
numerical study of a 10-parameter version of phenomenological MSSM (p10MSSM) and identify the ranges of values
of the axino DM mass and the reheating temperature that are consistent with data including large scale structure formation and big bang nucleosynthesis (BBN) constraints. We present the conclusions in Section \ref{sec:disc}. A number of technical issues are relegated to the Appendices. Appendix A summarizes the interaction Lagrangians of the Standard Model or MSSM fields with the axion and the axino. Appendix B presents some details of our calculation of TP of axinos and Appendix C describes some phase space integrals necessary to address the free-streaming length of axinos from NTP. 
In Appendix D, we describe the details of our numerical procedure.

%%%%%%%%%%%%%%%%%%%%%%%%%%%%%%%%%%%%%%%%%%%%%%%%%

\section{Axino with low reheating temperature -- general remarks}
\label{sec:axi1}

\subsection{Thermal production of axinos}

Our first aim is to understand the effects of low $T_R$ and of non-instantaneous reheating on axino abundance. 
Therefore, we will first discuss TP of axinos, 
focusing mainly on the KSVZ model. For concreteness, we will present the results obtained for the gluino and squark
masses of $m_{\tilde g}=m_{\tilde q}=1\,\mathrm{TeV}$, and we shall assume that the  $C_{aWW}=0$ (see Appendix A for details regarding parameters and
notations), 
unless indicated otherwise. We shall relax this assumption about the MSSM mass spectrum later.

In the high-$T_R$ regime, the scatterings associated with the $SU(3)_c$ group
dominate axino TP.\footnote{In literature, there exist different prescriptions for
treating the infrared divergence in relevant scattering cross-sections (see, e.g., \cite{Baer:2014eja}.
We conclude that there currently remains a factor of a few uncertainty in the thermal yield of axinos at high
$T_R$. In our numerical analysis, we will use the effective mass approximation\cite{Covi:2001nw}.
However,
as we will focus on the case of low $T_R$, this uncertainty will be of no consequence for
our main results; see  Section \ref{sec:axi2}. }
Were the reheating process instantaneous, we could identify the 
reheating temperature $T_R$ with the temperature $T_\mathrm{RD}$
at which the radiation dominated  epoch began.
% and the result (\ref{eq:ytp}) would hold.
However, with the RD epoch having been preceded by a reheating epoch,
during which the energy density of the oscillating and decaying inflaton dominated the Universe,
there is an additional contribution to $Y_{\tilde{a}}^\mathrm{TP}$
originating from a modified
relation $a(t)\sim T^{-8/3}$ between the scale factor of the Universe $a(t)$ and the temperature $T$.
A straightforward, but technical and rather involved calculation presented in Appendix B leads to the conclusion that this additional contribution is about 1/6 of the standard high $T_R$ result.
Loosely speaking, 
the existence of a reheating phase before the RD epoch effectively extends the period of relic production and available temperature range -- and therefore  the axino abundance increases.

\begin{figure}
\begin{center}
\includegraphics*[width=7cm]{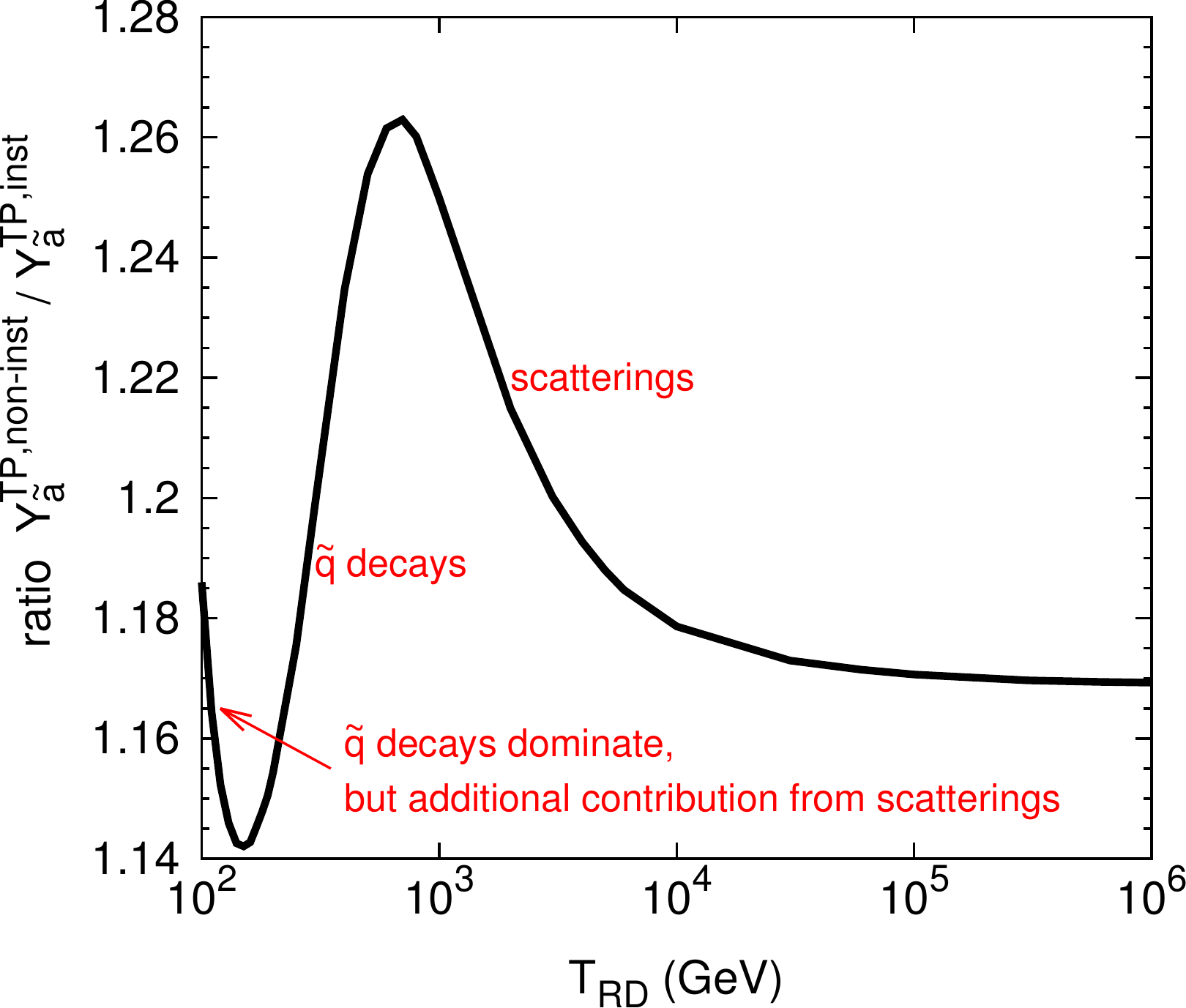}
\hspace{0.99cm}
\includegraphics*[width=7cm]{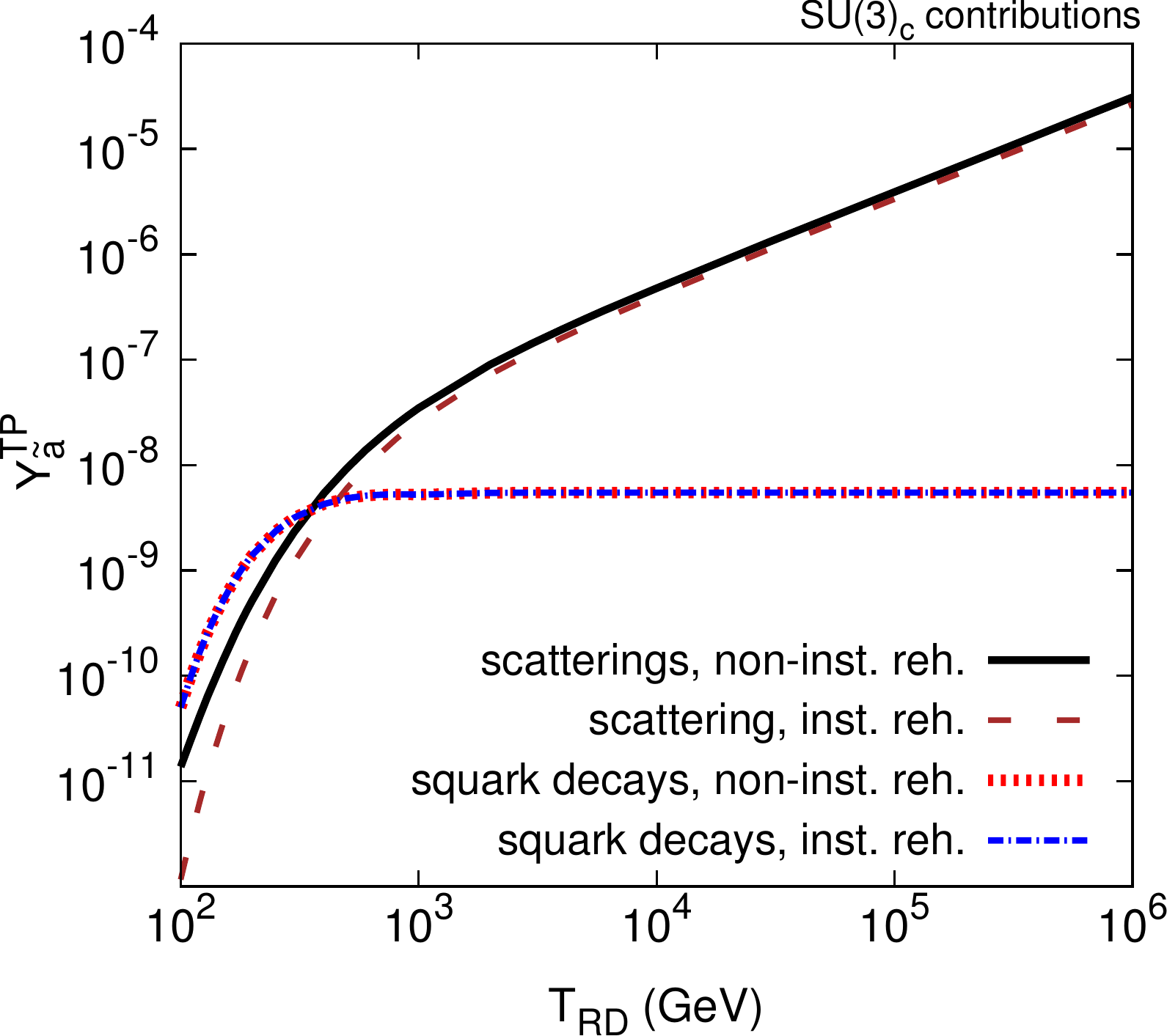}
\end{center}
\caption{
Left panel: The ratio of the axino abundances in the  KSVZ model obtained with the assumption of non-instantaneous ($Y_{\tilde{a}}^\mathrm{TP,non-inst}$) and instantaneous ($Y_{\tilde{a}}^\mathrm{TP,inst}$) reheating as a function of $T_{\textrm{RD}}$. The gluino and squark masses are set to $1\,\mathrm{TeV}$.
Right panel: Predictions for TP of axinos resulting from  scatterings and squark decays for $10^2\,\mathrm{GeV}<T_\mathrm{RD}<10^6\,\mathrm{GeV}$; results obtained for instantaneous  reheating are also shown.
\label{f1}
}
\end{figure}

One comment is in order here. In Ref.\ \cite{Strumia:2010aa}, which studied the same situation as described above,
a constant {\it reduction} of $Y_{\tilde a}^\mathrm{TP}$ by a factor of about 1/4 was reported. 
This apparent difference results from a different choice of a reference point.
In Ref.\ \cite{Strumia:2010aa} the results obtained for instantaneous and non-instantaneous reheating were compared for the same value of $T_R$, while we believe that it is more appropriate to compare both abundances at the same value of $T_\mathrm{RD}$,
which (in contrast to $T_R$ being a convenient shorthand notation for the inflaton decay rate) has a clear physical
meaning of the temperature that marks the beginning of the standard radiation dominated epoch.
As shown in Appendix B, we find an approximate relation $T_\mathrm{RD}\sim T_R/2$, so we can use these two temperatures
interchangeably when referring to orders of magnitude.

For intermediate values $10^2\,\mathrm{GeV}\simlt T_\mathrm{RD}\simlt 10^4\,\mathrm{GeV}$,
there is a phase space suppression of the scattering terms
associated with TeV-scale superparticles. 
For a given value of $T_\mathrm{RD}$ this effect is smaller in the case of non-instantaneous reheating, as the additional contribution to $Y_{\tilde a}^\mathrm{TP}$
stabilizes the result and the ratio of the abundances calculated for non-instantaneous and instantaneous reheating becomes slightly 
larger than $7/6\approx1.17$. 
This can be seen in the left panel of Fig.\ \ref{f1}.

\begin{figure}
\begin{center}
\includegraphics*[width=7cm]{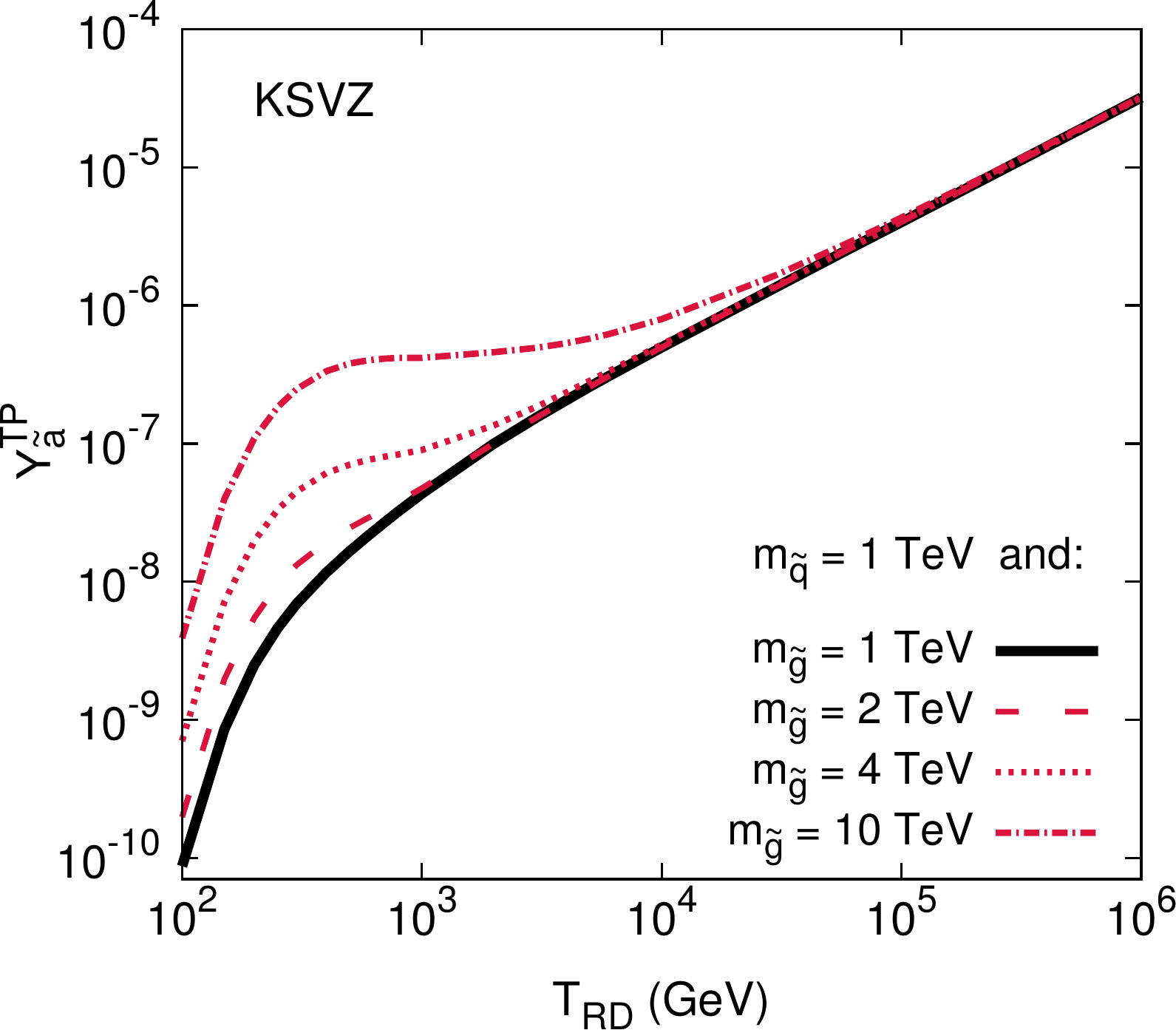}
\hspace{0.99cm}
\includegraphics*[width=7cm]{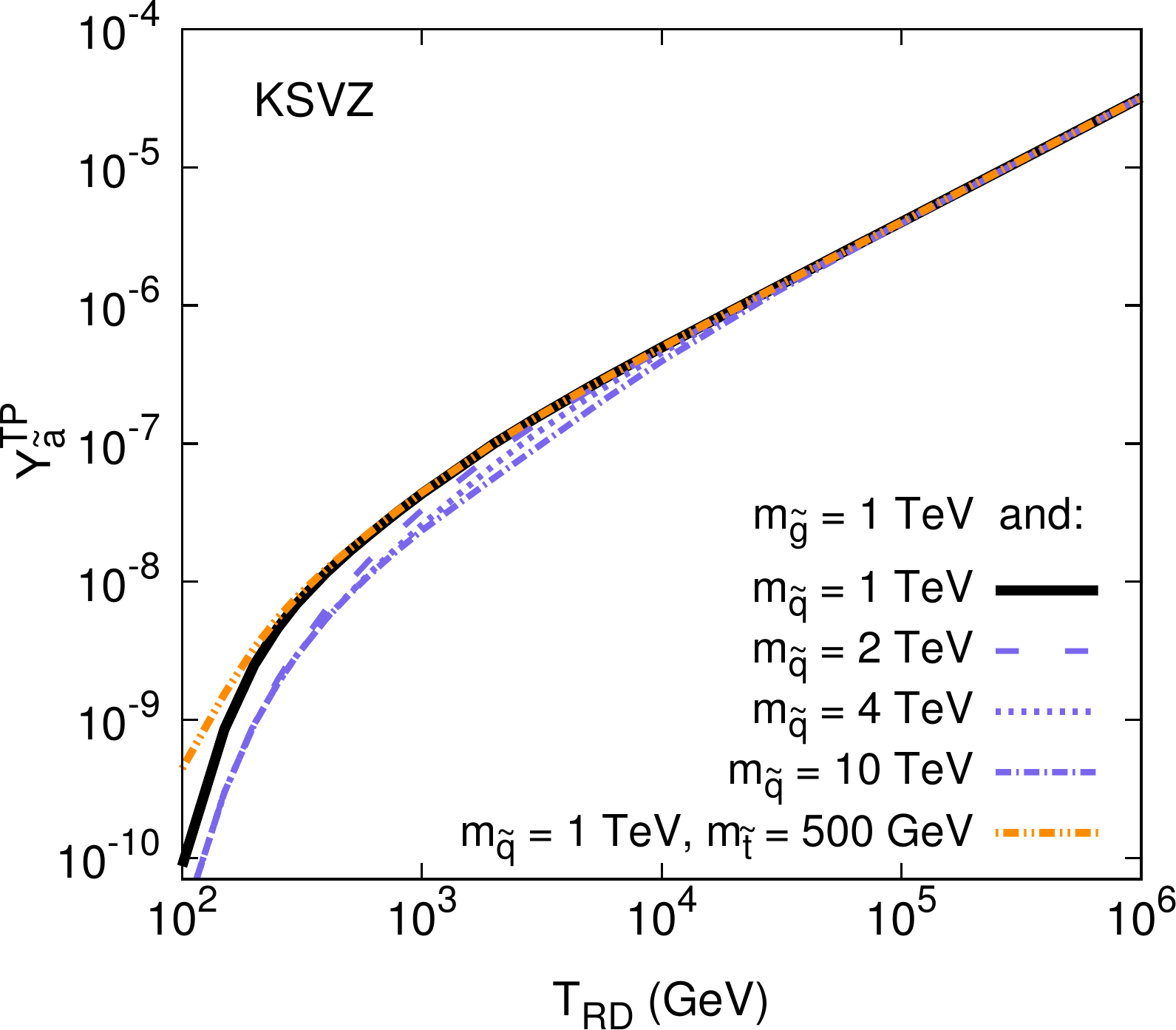}
\end{center}
\caption{Predictions for TP of axinos for different choices of the MSSM mass spectra, as indicated in the plots. Non-colour contribution is negligible here.
\label{f5}
}
\end{figure}

For $T_\mathrm{RD}$ smaller than the masses of strongly interacting particles (herein 1~TeV)
the ratio falls, as the contribution to axino TP from scatterings becomes small with respect to the contribution from 
squark and gluino decays.
This is because,
unlike scatterings, the decays 
depend rather weakly on the details of reheating as shown in the right panel of Fig.\ \ref{f1}. 
In principle, larger temperature values attainable with
non-instantaneous reheating may lead to a larger equilibrium number density of decaying squarks or gluinos and therefore
to an increase of $Y_{\tilde a}^\mathrm{TP}$, but this effect is  less important than the phase space suppression in the case of scatterings. 
Hence, as long as $Y_{\tilde a}^\mathrm{TP}$ is dominated by decays, the ratio in the left panel of Fig.\ \ref{f1} falls below 1.17. Fig.~\ref{f5} shows examples of the dependence of axino TP on the gluino and the squark masses.

The increase of the ratio of the abundances for the instantaneous and non-instantaneous reheating 
scenarios can be sizable only for $T_\mathrm{RD}\simlt 100\,\mathrm{GeV}$,
if the parameter $C_{aYY}$ parametrizing the coupling between the axino and the $U(1)_Y$
gauge bosons and gauginos is equal to zero or if bino-like neutralino is heavier than about $500$\,GeV.
Otherwise, axino TP is at low temperatures dominated by the decays of the light bino which is again rather insensitive to
the details of reheating if $T_\mathrm{RD}$ is comparable to the bino mass.

However, for such low values of $T_\mathrm{RD}$, even for vanishing $C_{aYY}$ and/or heavy bino,
TP often gives an abundance which is a very small fraction of that required for DM, so details of reheating are
irrelevant in that case. One should also remember that for very low values of $T_\mathrm{RD}$ axinos
are decoupled from kinetic but not from chemical equilibrium and their distribution function can differ from
that for the equilibrium case \cite{Monteux:2015qqa}, but for axinos this only happens for the DM abundance
dominated by NTP, so this effect has no consequences for our study.

We can see that the main effect of non-instantaneous reheating on the axino TP originates from a modified contribution from scattering.
Therefore, the details of reheating play a minor role in DFSZ models, as in this case TP is typically dominated by thermal higgsino decays \cite{Brandenburg:2004du,Chun:2011zd}.

\subsection{Non-thermal production of axinos}

\begin{figure}
\begin{center}
\includegraphics*[width=8cm]{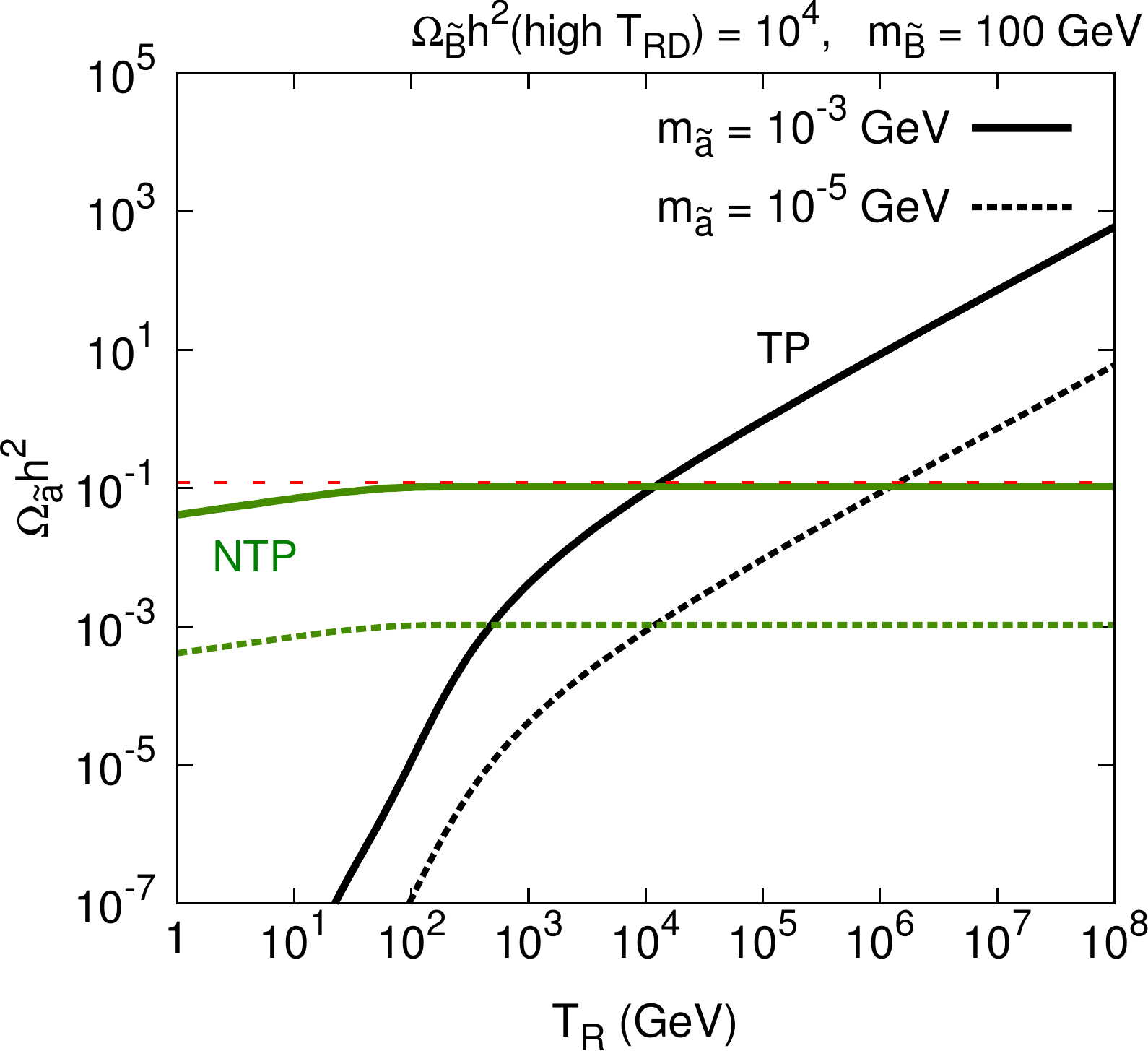}
\hspace{0.5cm}
\includegraphics*[width=8cm]{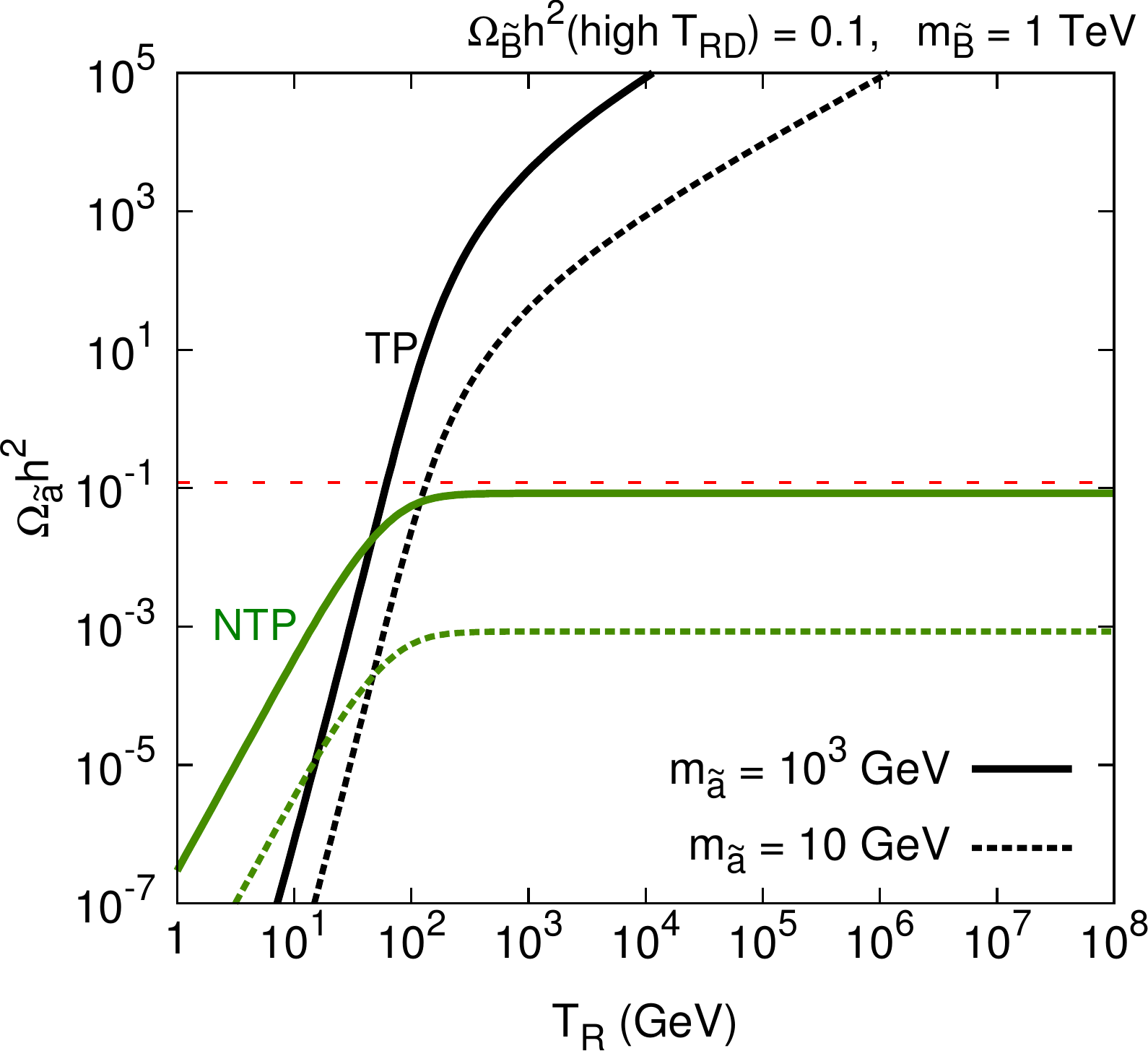}
\end{center}
\caption{Cosmological parameters $\Omega_{\tilde a} h^2$ resulting from TP and NTP of KSVZ axinos for different values of the reheating temperature $T_R$. Gluino and squark masses are set to 1~TeV, while the bino masses was set to $100\,\mathrm{GeV}$ with the rest of the MSSM spectrum arranged so that the high-$T_R$
bino abundance is $\Omega_{\widetilde B}h^2=10^4$ (left panel) and to $1\,\mathrm{TeV}$ with the rest of the MSSM spectrum arranged so that the high-$T_R$
bino abundance is $\Omega_{\widetilde B}h^2=0.1$ (right panel).
\label{fNTP}
}
\end{figure}

For low enough values of $T_{\textrm{RD}}$ axino TP is highly suppressed and it is the non-thermal contribution to the axino relic density that dominates. However, NTP is also affected by an additional entropy production during the reheating period provided that $T_{\textrm{RD}}$ is low enough so that the LOSPs freeze out before the RD epoch (see, e.g., a recent discussion about similar issue in the case of gravitino DM~\cite{Roszkowski:2014lga}).\footnote{The most conservative lower bounds on the reheating temperature from BBN can be derived in terms of a successful neutrino thermalization for electromagnetic energy emission, $T_R > 0.7\,$MeV and
$T_R >  4 - 5\,$MeV for weak-scale parent particles in terms of $n-p$ conversion due to
hadron emissions\cite{Kawasaki:1999na,Kawasaki:2000en}.} This results in a suppression of $\Omega_{\textrm{LOSP}}h^2$ below 
the value
$\Omega_{\textrm{LOSP}}h^2(\textrm{high }T_{\textrm{RD}})$
obtained 
in the standard cosmological scenario where the LOSP freeze-out occurs in the RD epoch.
Consequently, for a fixed $m_{\tilde{a}}/m_{\textrm{LOSP}}$ ratio  in Eq.~(\ref{eq:Oh2ntp}), $\Omega^{\textrm{NTP}}_{\tilde{a}}h^2$ also decreases. The lower the reheating temperature, the longer  the period between the LOSP freeze-out and the beginning of the RD epoch, so  the LOSPs are effectively diluted. As a result  $\Omega^{\textrm{NTP}}_{\tilde{a}}h^2$ decreases with $T_{\textrm{RD}}$. This is illustrated in Fig.~\ref{fNTP} where we show both TP and NTP contributions to $\Omega_{\tilde{a}}h^2$ for two selected SUSY spectra with bino LOSP mass equal to $100\,$GeV and $\sim 1\,$TeV, while squark and gluino masses are set to $1\,$TeV. The rest of the SUSY spectrum is in both cases chosen such as to obtain LOSP yield at freeze-out corresponding to $\Omega_{\tilde{B}}h^2(\textrm{high }T_{\textrm{R}}) = 0.1$ or $10^4$.

\subsection{BBN and WDM constraints}

The BBN constraints for the axino can be analyzed similarly to those for the 
gravitino (see e.g.\ \cite{Roszkowski:2004jd,Cerdeno:2005eu,Steffen:2006,Kanzaki:2006,Pospelov:2006sc,Jedamzik:2007qk,Kawasaki:2007xb,Kawasaki:2008qe});
they are typically mild. This is because the lifetime of the  LOSP decaying to the axino usually hardly exceeds 0.1~sec., unless one considers very light LOSP, allows for a very strong mass degeneracy, $m_{\tilde a}\approx m_\mathrm{LOSP}$
or considers a LOSP whose 2-body decays to axinos are forbidden (i.e.\ bino with $C_{aYY}=0$). 
However, if the neutralinos are too light, they have a very small relic abundance, which is further suppressed by a small value of $T_R$,
and there are simply too few decaying LOSPs around to put primordial nucleosynthesis in danger (in this case, the correct axino DM abundance must originate from TP). The only exception is a scenario with a very light axino and a very light bino LOSP with extremely small annihilation rate, suppressed by large slepton masses. In this case, the relic abundance of bino LOSPs can be very large and the correct axino DM relic density is obtained from NTP by a suppression with a small axino mass.

A more dramatic change appears when we set $C_{aYY}=0$, thereby assuming that the axino interacts only with the $SU(3)_c$ gauge sector.
For bino LOSP, the 2-body bino decay into axino and a photon or $Z$ boson are then disallowed and the dominant bino decay channel $\widetilde B\to q\bar{q}\tilde{a}$ involves
an effective axino-quark-squark interaction vertex discussed in \cite{Covi:2002vw}. Applying the formulae obtained therein,
we can estimate the bino lifetime as (see also Appendix C)
\beq
\tau_{\widetilde B} \approx 360\,\mathrm{sec}\,\left( \frac{100\,\mathrm{GeV}}{m_{\widetilde B}}\right)^5 \left( \frac{m_{\tilde q}}{1\,\mathrm{TeV}}\right)^4\left( \frac{1\,\mathrm{TeV}}{m_{\tilde g}}\right)^2 \left( \frac{f_{\tilde a}}{10^{11}\,\mathrm{GeV}}\right)^2 \, .
\label{eq:cayy0}
\eeq
With a longer LOSP lifetime, the BBN constraints are significantly more severe.

For small enough values of $m_{\tilde a}$, axinos may become warm dark matter (WDM) 
with a free streaming length which is too large to account for observed structures in the Universe.\footnote{For alternative, cosmologically viable scenario with axino WDM from late-time saxion decays see\cite{Choi:2012zna}.}
Thermally and non-thermally produced axinos have different rms velocities and hence these constraints
from structure formation depends on the fraction of the WDM axinos in DM (for a discussion about non-thermally produced WDM see, e.g.,\cite{Hisano:2000dz,Jedamzik:2005sx}).
We use the 95\%~CL exclusion limits from Ref.\ \cite{Boyarsky:2008xj}. 
For TP domination these limits translate to $m_{\tilde{a}}\simlt 5\,\mathrm{keV}$, while 
the case of NTP domination and the mixed case require a more careful analysis 
In particular,  for $C_{aYY}=0$ the calculation of the rms axino velocity requires an integration over 3-\ and 4-particle phase space (bino and stau LOSPs, respectively); we provide technical details of this computation in Appendix C.

\section{Axino with low reheating temperature in the MSSM}
\label{sec:axi2}

Having discussed the predictions for TP and NTP of axinos for different values of the reheating temperature $T_R$
and for different patterns of the MSSM mass spectra, we are now ready to present and discuss the results
of our extensive numerical scan. There are shown in Fig.~\ref{fscan} for $f_a=5\times 10^{9}\,\mathrm{GeV}$ and $10^{11}\,\mathrm{GeV}$.
In comparison with the results of previous studies, we find a number of differences in the shape of the allowed region of the axino mass
and the reheating temperature. We discuss them separately for the bino LOSP and for the wino and higgsino LOSP.

\subsection{Bino LOSP}
\label{sec:axi2-bino}

Firstly,  similarly to Ref.~\cite{Choi:2011yf} we find an upper bound on $T_R$ and a lower
bound on $m_{\tilde a}$, beyond which axino DM becomes too warm.  For $T_R\simlt10^6\,\mathrm{GeV}$ the region allowed 
by the BBN or WDM constraints extends to smaller values of $m_{\tilde a}\sim10^{-4}\,\mathrm{GeV}$ than in previous analyses. For $T_R\simlt10^3\,\mathrm{GeV}$ these points correspond to dominant NTP from a very large relic density of bino LOSP with a very small annihilation cross-section dominated by a $t$-channel exchange of multi-TeV sleptons. Such points, are however excluded by both BBN constraints 
(bino LOSP has a long lifetime and a sizable hadronic branching fraction, cf.\ \cite{Covi:2009}) 
and by WDM constraints (with dominant NTP, one needs $m_{\tilde a}>30\,\mathrm{MeV}$). Eventually,
the lower bounds for $m_{\tilde a}$ at fixed $T_R$ are similar to those obtained in \cite{Choi:2011yf}, but in our analysis they result from incorporating additional cosmological constraints. 

\begin{figure}
\begin{center}
\includegraphics*[width=7cm]{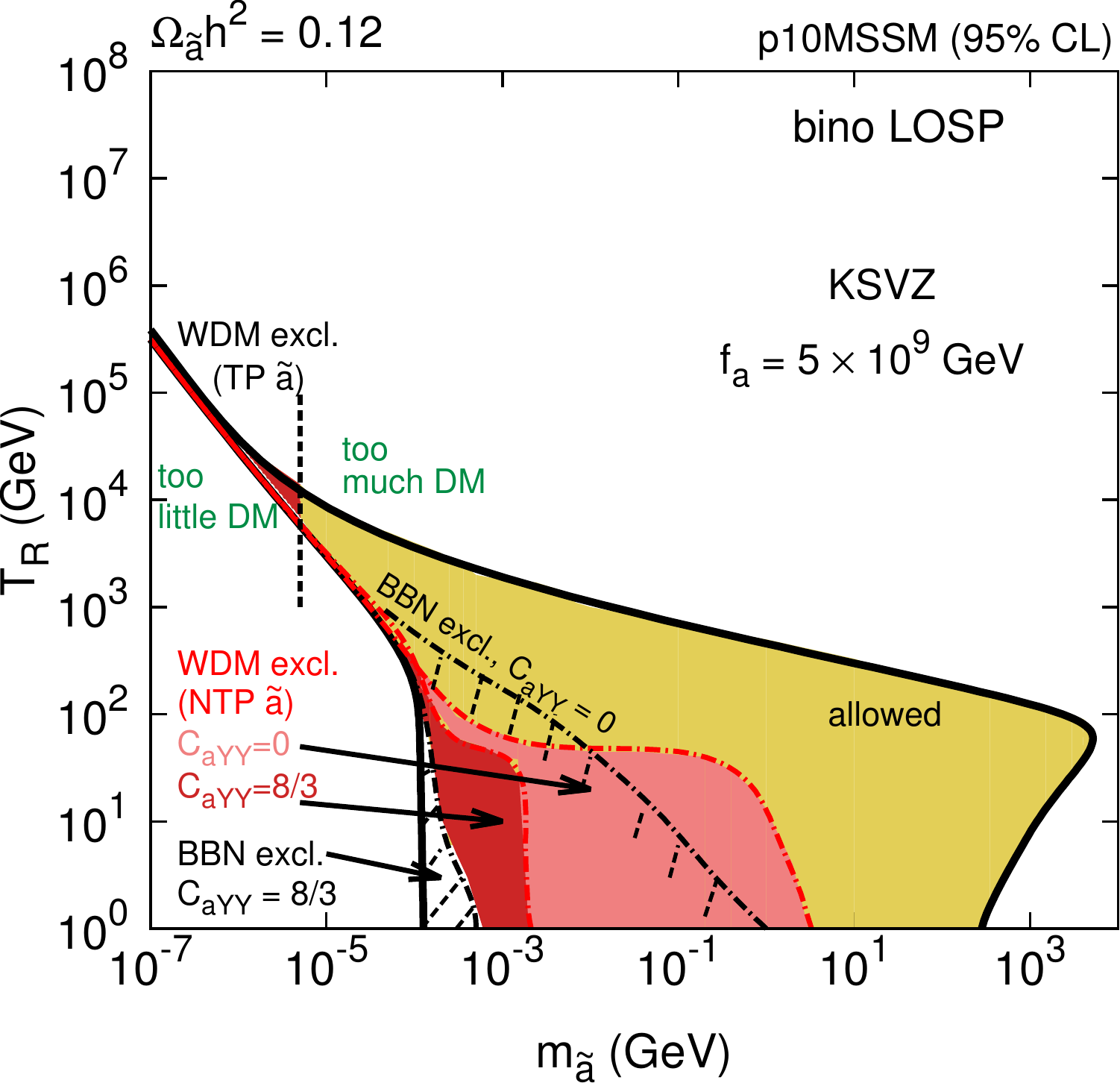}
\hspace{0.99cm}
\includegraphics*[width=7cm]{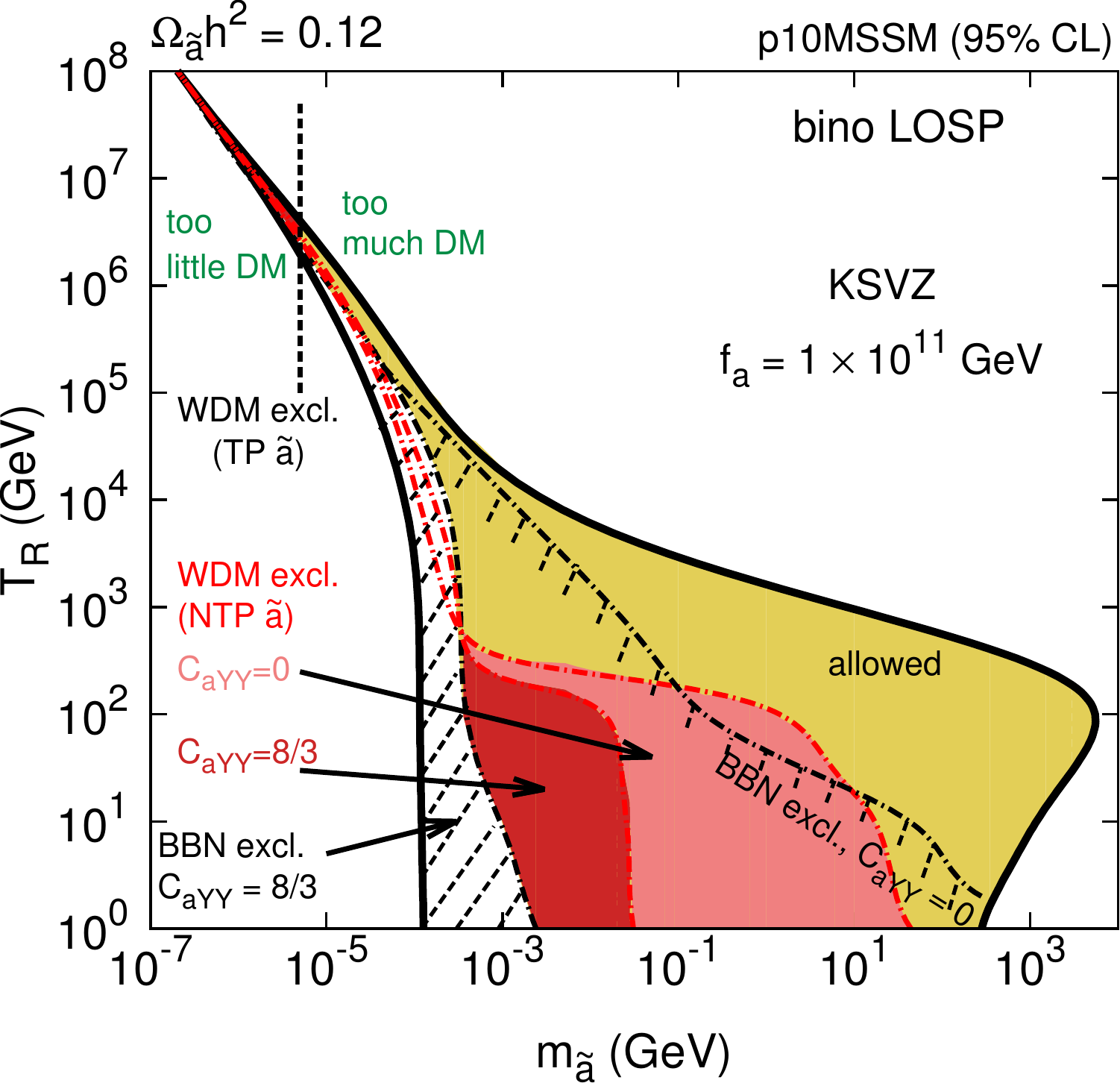}
\end{center}
\caption{Results of the numerical scan for bino LOSP projected onto the $(m_{\tilde a},T_R)$ plane for the KSVZ axino with $f_a = 5\times 10^9$ (left panel) and $f_a = 1\times 10^{11}$ (right panel). The area bounded by thick solid (partially shaded) lines denotes acceptable regions for which MSSM parameters consistent with all the bounds listed in Appendix D, including DM energy density, could be found. 
Inside these regions, we marked in red the regions excluded by axino dark matter being too warm (WDM constraints). 
Dashed vertical lines denote the lower bounds on the axino mass coming from WDM constraints for TP only. Regions excluded by BBN constraints are either dashed ($C_{aYY}=8/3$) or marked with a dash-dotted lines ($C_{aYY}$=0).
\label{fscan}
}
\end{figure}

\begin{figure}
\begin{center}
\includegraphics*[width=7cm]{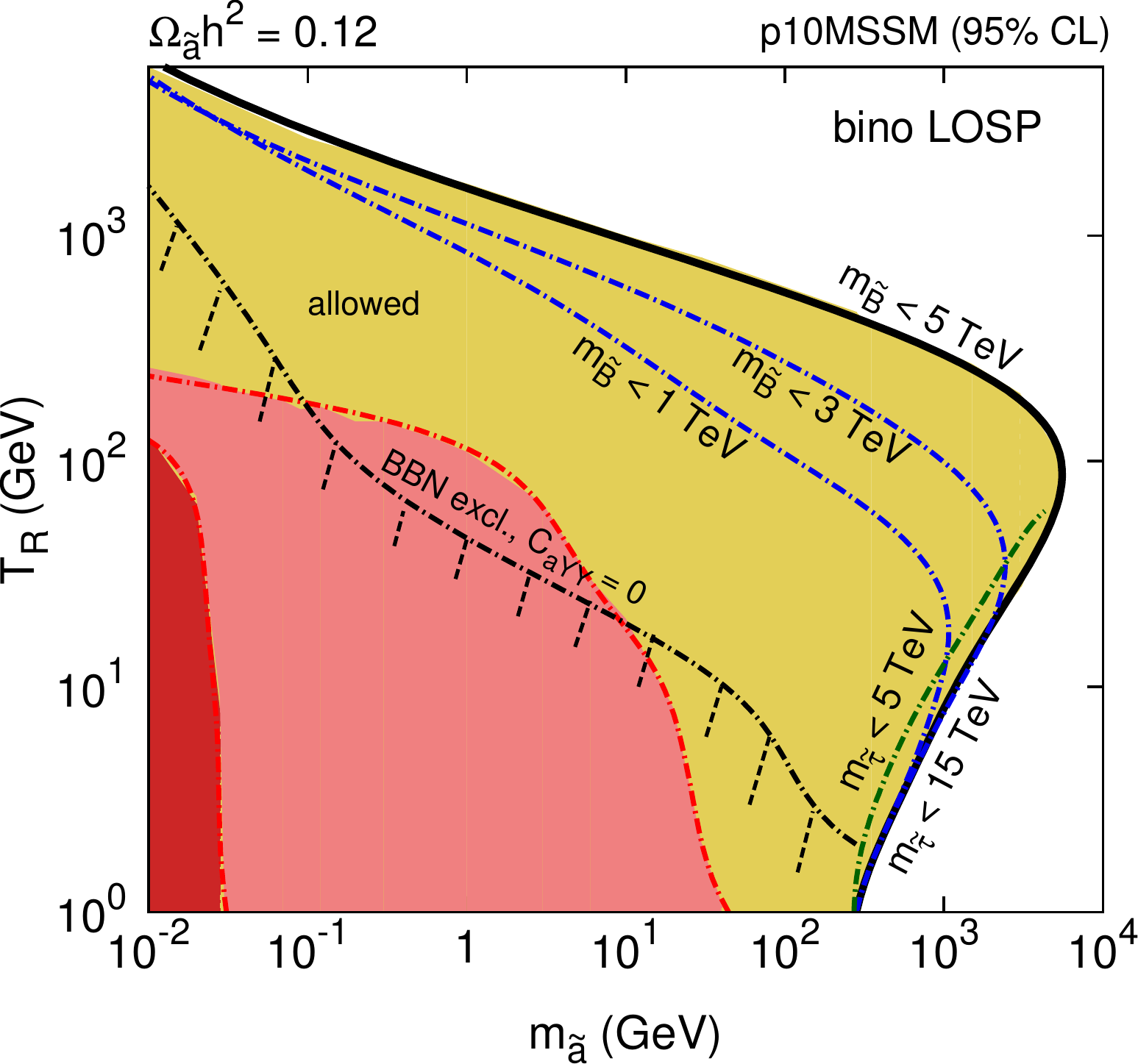}
\hspace{0.99cm}
\includegraphics*[width=7cm]{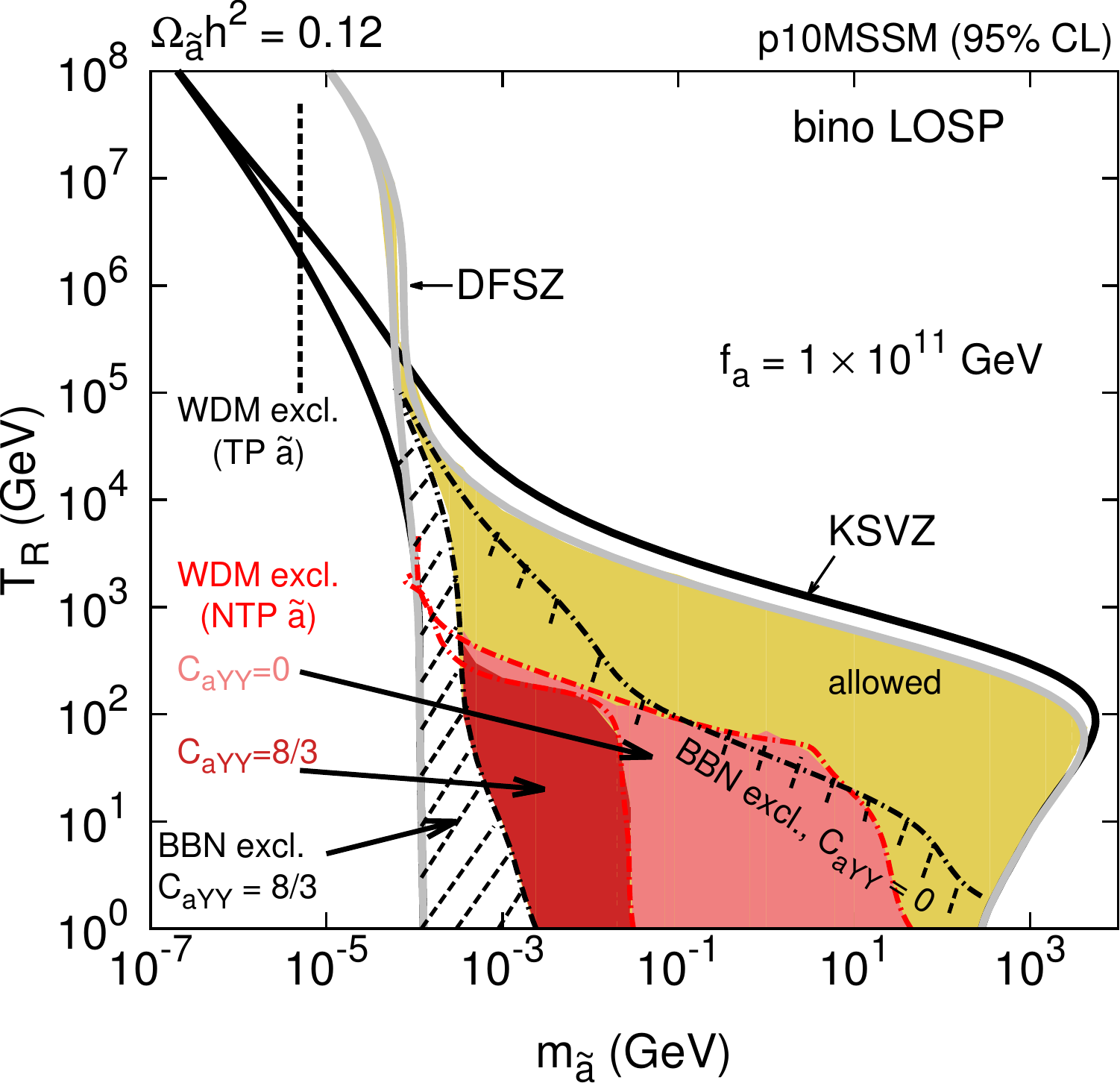}
\end{center}
\caption{%
The left panel is a blow-up of the bottom-right corner of the right panel of Fig.~\ref{fscan}; boundaries of the allowed region corresponding to different ranges of the bino mass, scanned up to 1, 3 and 5~TeV, and different ranges of the stau mass, scanned up to 5 and 15~TeV are shown (cf.~Table~\ref{tabp10MSSM} for mass range dependencies). The right panel shows the results for the DFSZ model with $f_a=1\times10^{11}\,\mathrm{GeV}$; for comparison the contours corresponding to the allowed region in the KSVZ model (right panel of Fig.\ \ref{fscan}) are
also shown.
\label{fscan2}
}
\end{figure}

Secondly, looking at the rightmost part of each plot in Fig.\ \ref{fscan}, we see an upper bound on $m_{\tilde a}$. 
In this respect our result visually resembles that of Ref.~\cite{Choi:2011yf}, but this similarity follows from very different
physical assumptions. In Ref.~\cite{Choi:2011yf} three typical choices of fixed axino abundance from NTP, $Y_{\tilde a}^\mathrm{NTP}$,
were considered, which led to upper bounds on axino mass for $Y_{\tilde a}^\mathrm{NTP}>0$. When NTP was the dominant source of axino DM, these bounds on $m_{\tilde a}$ did not depend on $T_R$. In contrast, in our analysis we obviously do not make such a restrictive assumption and the maximum value of $m_{\tilde a}$ that we obtain is related to the maximum value of the LOSP mass.
More precisely,
the largest value of $m_{\tilde a}$ is obtained for $T_R\sim\mathcal{O}(10^2\,\mathrm{GeV})$,
corresponding to the freeze-out temperature of the bino LOSP. 
This maximum $m_{\tilde a}$ corresponds
to the largest available values of the bino LOSP mass and the largest available slepton masses (cf.~Table~\ref{tabp10MSSM}). These two features have the same effect: one expects a larger relic density for heavy particles and there is also a big suppression of the annihilation cross-section due to large masses of the intermediate particles. We show these aspects of our results in the left panel of 
Fig.~\ref{fscan2}, where we show how the allowed region changes with different assumptions about the largest possible bino and slepton masses. 

Furthermore, we see a decrease of the maximum allowed $m_{\tilde a}$ for $T_{R}$ falling below $\mathcal{O}(10^2 \,\mathrm{GeV})$.
This can be understood taking into account that during reheating
there is  entropy production due to inflaton decays. As a consequence, 
the LOSP relic density becomes suppressed,
if $T_R$ falls significantly below the freeze-out temperature.
For a given value of $T_R$ 
the suppression is stronger 
for binos with larger masses.
This confines the allowed region to smaller values of bino mass and, as $m_{\tilde a}<m_{\widetilde B}$, to smaller values 
of axino mass.

%%%

We also note that, for $T_R\simlt 10^4\,\mathrm{GeV}$ the calculations for TP of axinos cannot be fully trusted, as the $SU(3)_c$ gauge coupling becomes large (see \cite{Choi:2013} and references therein). This poses no problem for $T_R\simlt 10^2\,\mathrm{GeV}$,
as NTP of axinos is then dominant, but in the window $10^2\,\mathrm{GeV}\simlt T_R\simlt 10^4\,\mathrm{GeV}$ the upper bounds on $m_{\tilde a}$ or $T_R$ should be treated as approximate.
Changing $f_{\tilde a}$ mostly leads to a shift of the allowed region in the $(m_{\tilde a},T_{R})$ plane, as shown in the right panel of
Fig.~\ref{fscan}. 

The difference between regions of the $(m_{\tilde a},T_R)$ plane allowed in the KSVZ and DFSZ models is presented in
the right panel of Fig.\ \ref{fscan2}. For large values of $T_R\simgt 10^5\,\mathrm{GeV}$ it can be
traced back to the fact that $\Omega^\mathrm{TP}_{\tilde a}h^2$ scales as $m_{\tilde a}T_R$ and $m_{\tilde a}$
for the KSVZ and DSFZ models, respectively  \cite{Brandenburg:2004du,Chun:2011zd}. 
For smaller values of $T_R$ the bulk of the allowed region is very similar
for both models. For a fixed $m_{\tilde a}\simgt 1\,\mathrm{MeV}$, there is still a small difference between the largest allowed $T_R$
corresponding to the TP dominance. This results from different sources of thermal contributions to axino DM from decays. In KSVZ models the most important decays are those of colored particles; these are loop-suppressed with respect to decays of neutralinos with a non-negligible higgsino fraction in DFSZ models. Therefore, in DFSZ models TP of axinos from decays is much more efficient which may lead to overproduction of axino DM for values of $m_{\tilde a}$ and $T_R$ corresponding to the correct axino DM density within the KSVZ scheme.

%%%%%%%%%%%%%%%%%%%%%%%%%%%%%%%%%%%%%%%%%%%%%%%%%

\subsection{Wino, higgsino and stau LOSP}
\label{sec:axi2-whino}

\begin{figure}
\begin{center}
\includegraphics*[width=7cm]{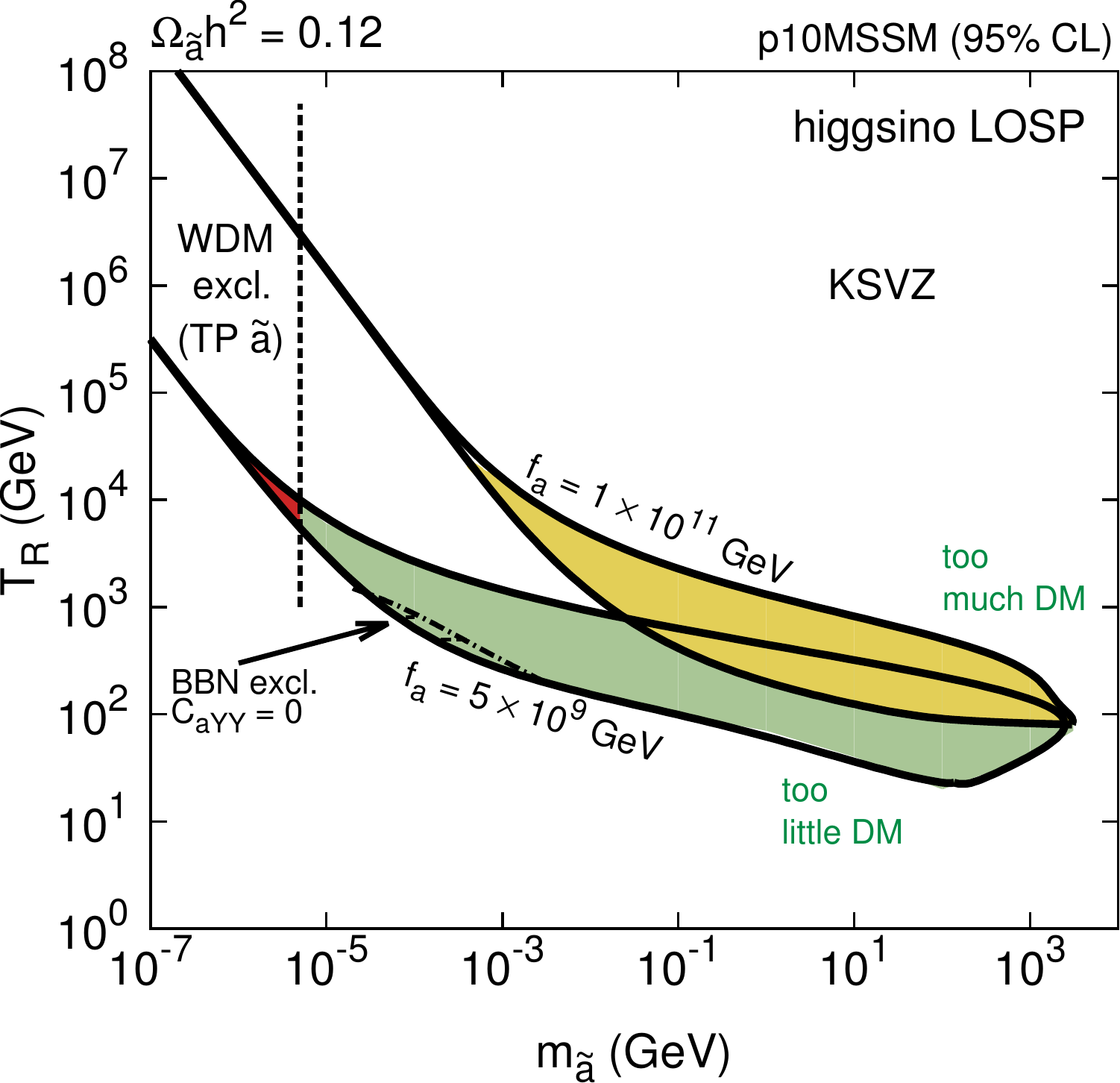}
\hspace{0.99cm}
\includegraphics*[width=7cm]{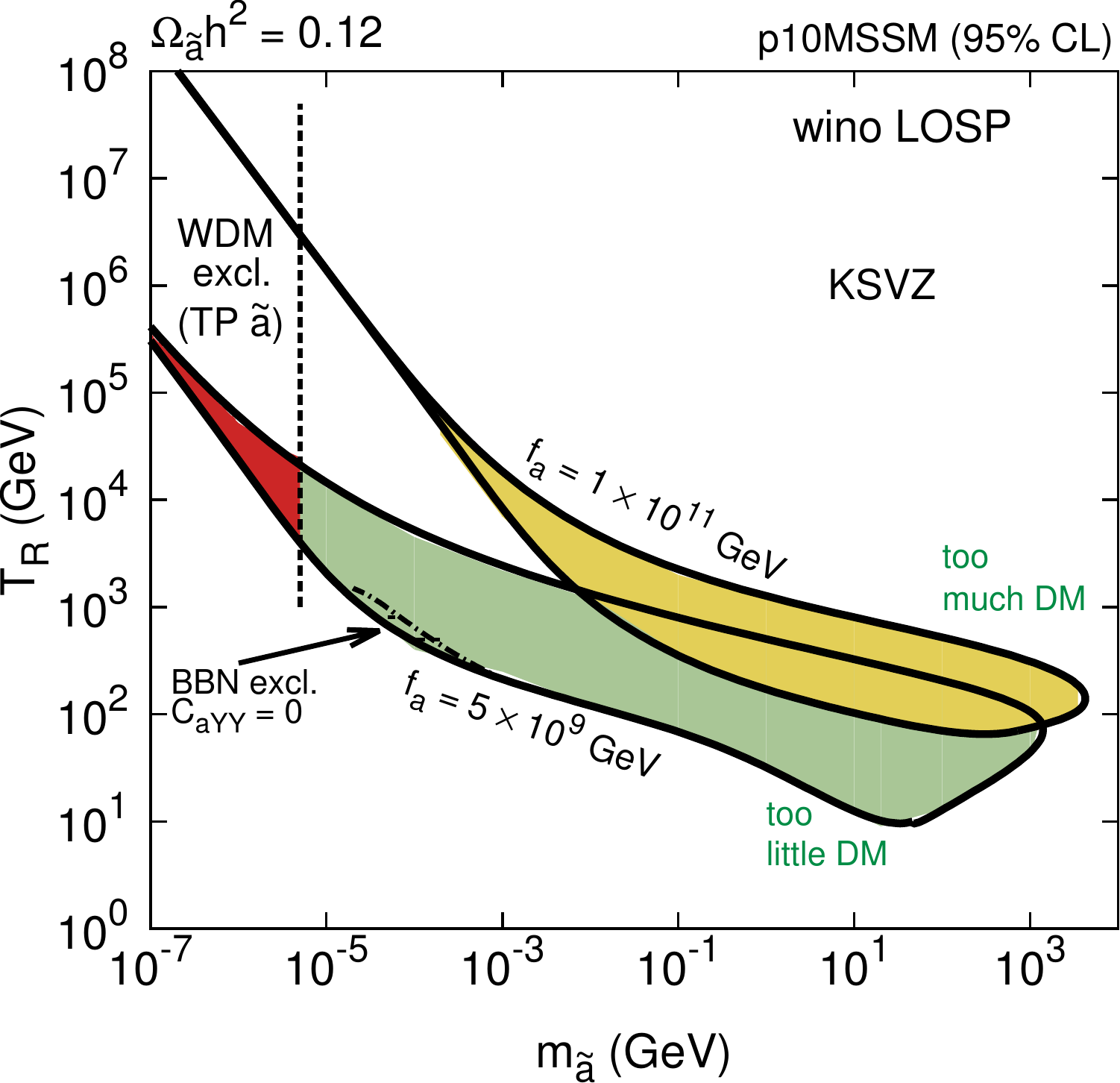}
\end{center}
\caption{%
The same as in Fig.\ \ref{fscan}, but for higgsino LOSP (left panel) and for wino LOSP (right panel).
\label{fhigwin}
}
\end{figure}

\begin{figure}
\begin{center}
\includegraphics*[width=7cm]{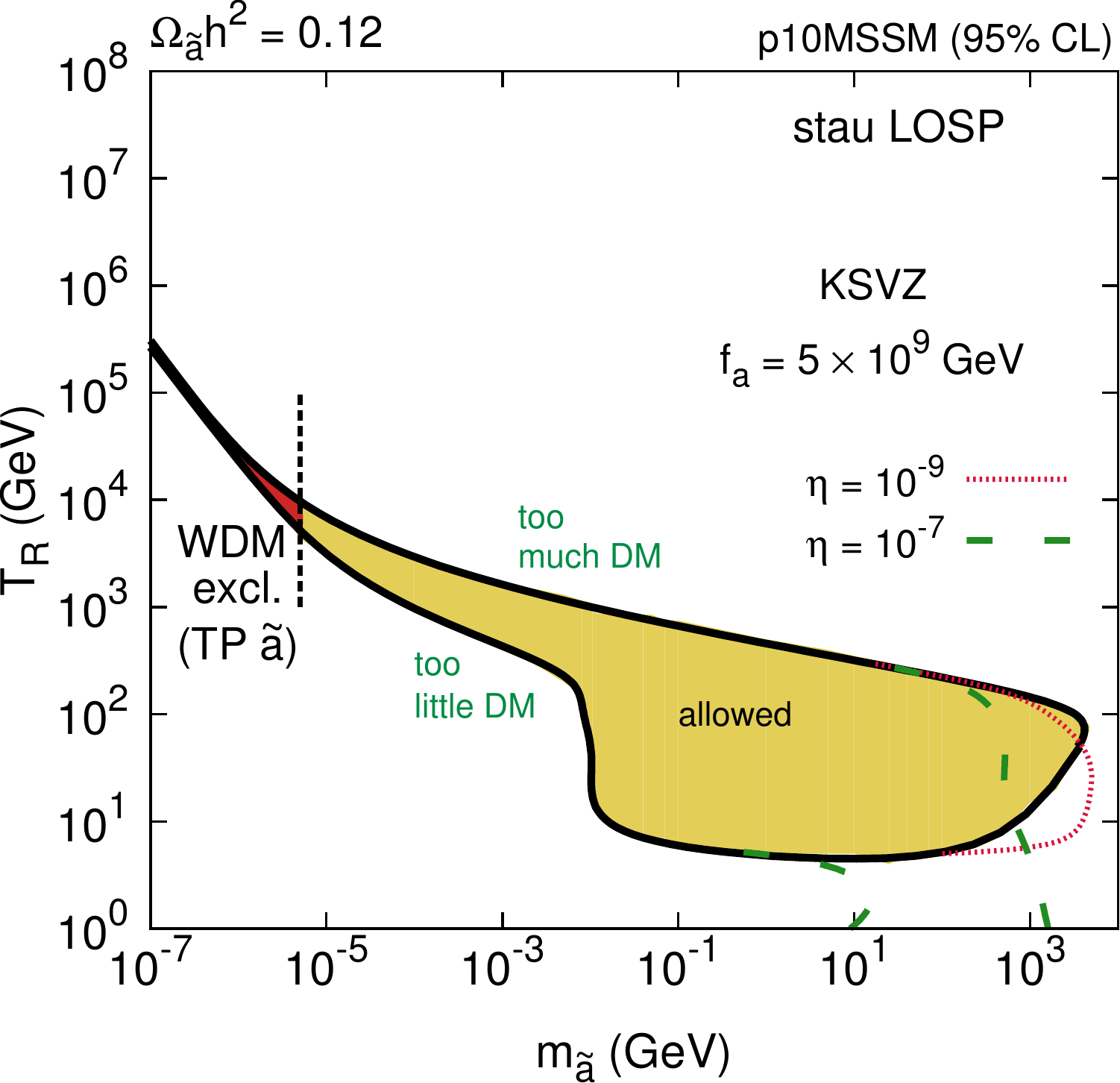}
\hspace{0.99cm}
\includegraphics*[width=7cm]{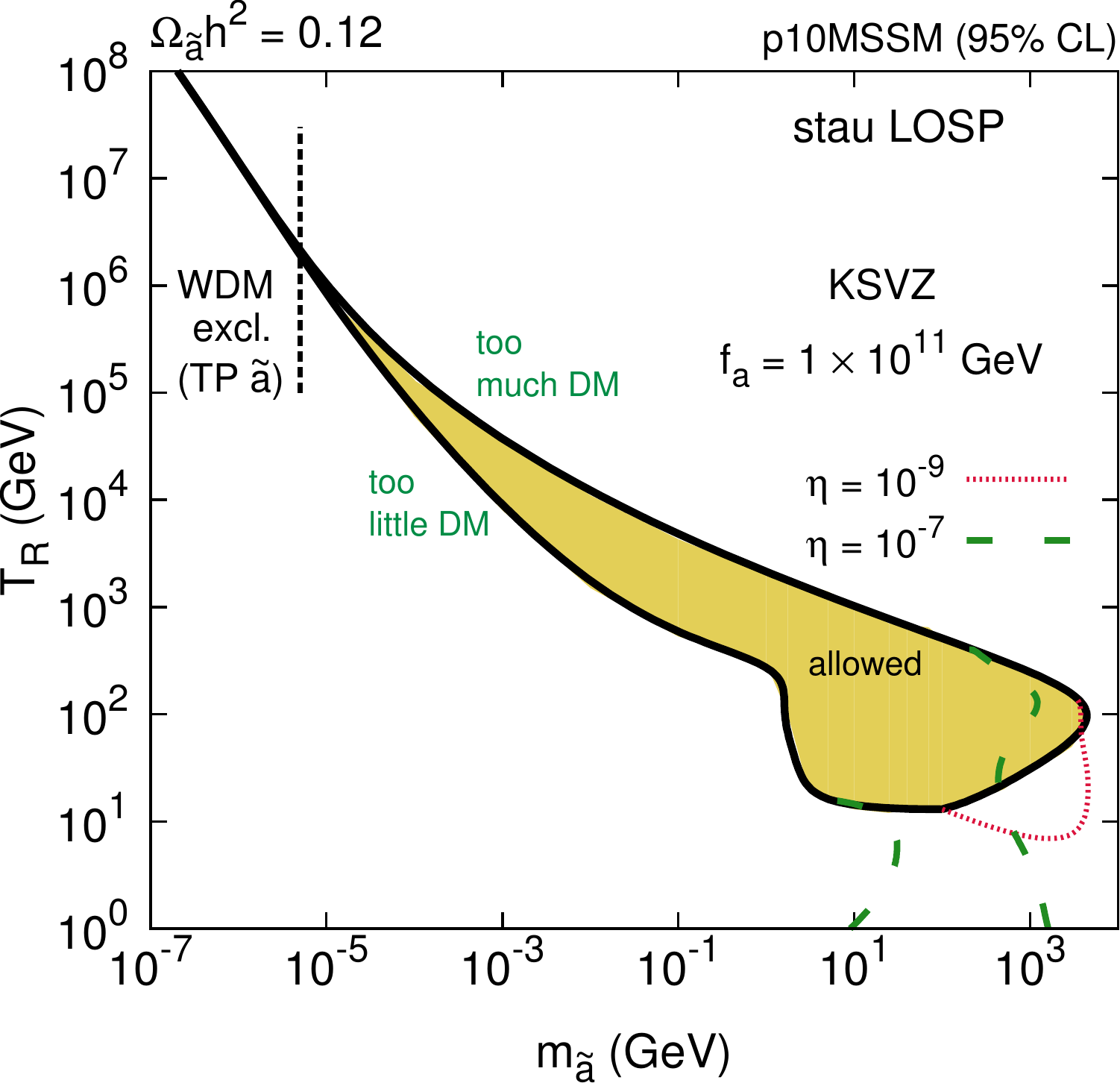}
\end{center}
\caption{%
The same as in Fig.\ \ref{fscan}, but for stau LOSP. Additionally, regions allowed with direct and cascade decays of the inflaton field (see Section \ref{sec:direct}) are shown.
\label{fstau}
}
\end{figure}

In Fig.\ \ref{fhigwin} we show the allowed regions of the $(m_{\tilde a},T_R)$ plane for the wino and higgsino LOSP
and in Fig.\ \ref{fstau} the same for the stau LOSP,
with constraints imposed in the same way as described in Section \ref{sec:axi2-bino}. We again show the results for two values
of $f_a=5\times 10^9\,\mathrm{GeV}$ and $10^{11}\,\mathrm{GeV}$. We restrict our analysis to the KSVZ model, since we have seen that
for low $T_R$ the predictions of the KSVZ an DFSZ models do not differ very much.

The main difference with respect to the bino LOSP case (Fig. \ref{fscan}) is that $T_R$ is now bounded from below. This is because for winos, higgsinos and staus (unlike for binos)
the masses of the states
mediating annihilations
cannot be much larger than the masses of annihilating particles. 
Hence, the annihilation cross-section of these particles cannot be made very small by increasing the mass of the intermediate states. Nonetheless, by varying MSSM parameters, we find examples of axino DM and the higgsino or wino or stau LOSP in which the observed DM abundance originates mainly from TP (upper parts of the allowed regions) or NTP (lower parts of the allowed regions). This may seem at odds with a recent analysis of axino DM in the DFSZ model with the higgsino LOSP \cite{Co:2015pka}
which studied both TP (freeze-in of axinos) and NTP (LOSP freeze-out), and concluded that TP always dominates. This apparent difference
can be traced to the fact that in Ref.\  \cite{Co:2015pka} two examples of the MSSM mass spectrum are studied, while our numerical
analysis extends to larger values of the masses of the supersymmetric particles and thus allows larger LOSP relic abundances. This in turn can give rise to the observed axino DM abundance {\em via} NTP even with additional entropy production from inflaton decays. 

In the case of the stau LOSP one can obtain significant contribution to the axino relic density from TP also for low values of $T_R\sim 10\,$GeV. It is due to decays of light stau LOSPs being still in thermal equilibrium. The mass of the axino is then typically significantly lower than $m_{\tilde{\tau}_1}$, which suppresses the NTP contribution. However, one needs to take into account that such a scenario is constrained by the LHC searches for direct production of staus with missing energy\cite{Aad:2014yka}. When treating this we calculate the relevant production cross section with \madgr\cite{Alwall:2014hca}.

\subsection{Direct and cascade decays of the inflaton field}
\label{sec:direct}

In our analysis so far, we have made an assumption that there are no direct and cascade decays of the inflaton field to axino DM particles.
Here, we would like to study the impact of such decays on the allowed ranges of axino mass and reheating temperature,
following the model-independent approach used in \cite{Gelmini:2006Feb,Gelmini:2006May}.

\begin{figure}
\begin{center}
\includegraphics*[width=7cm]{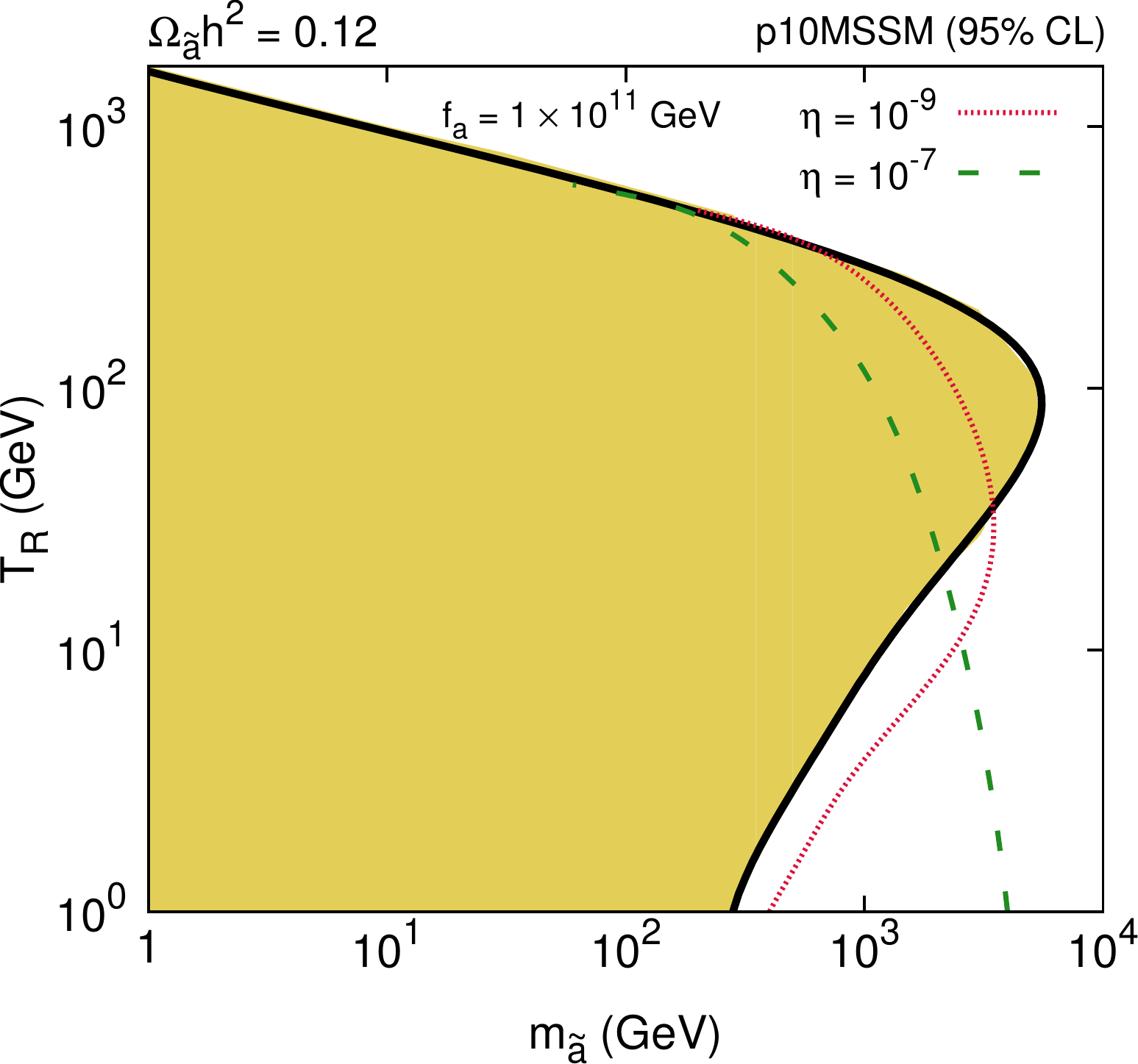}
\hspace{0.99cm}
\includegraphics*[width=7cm]{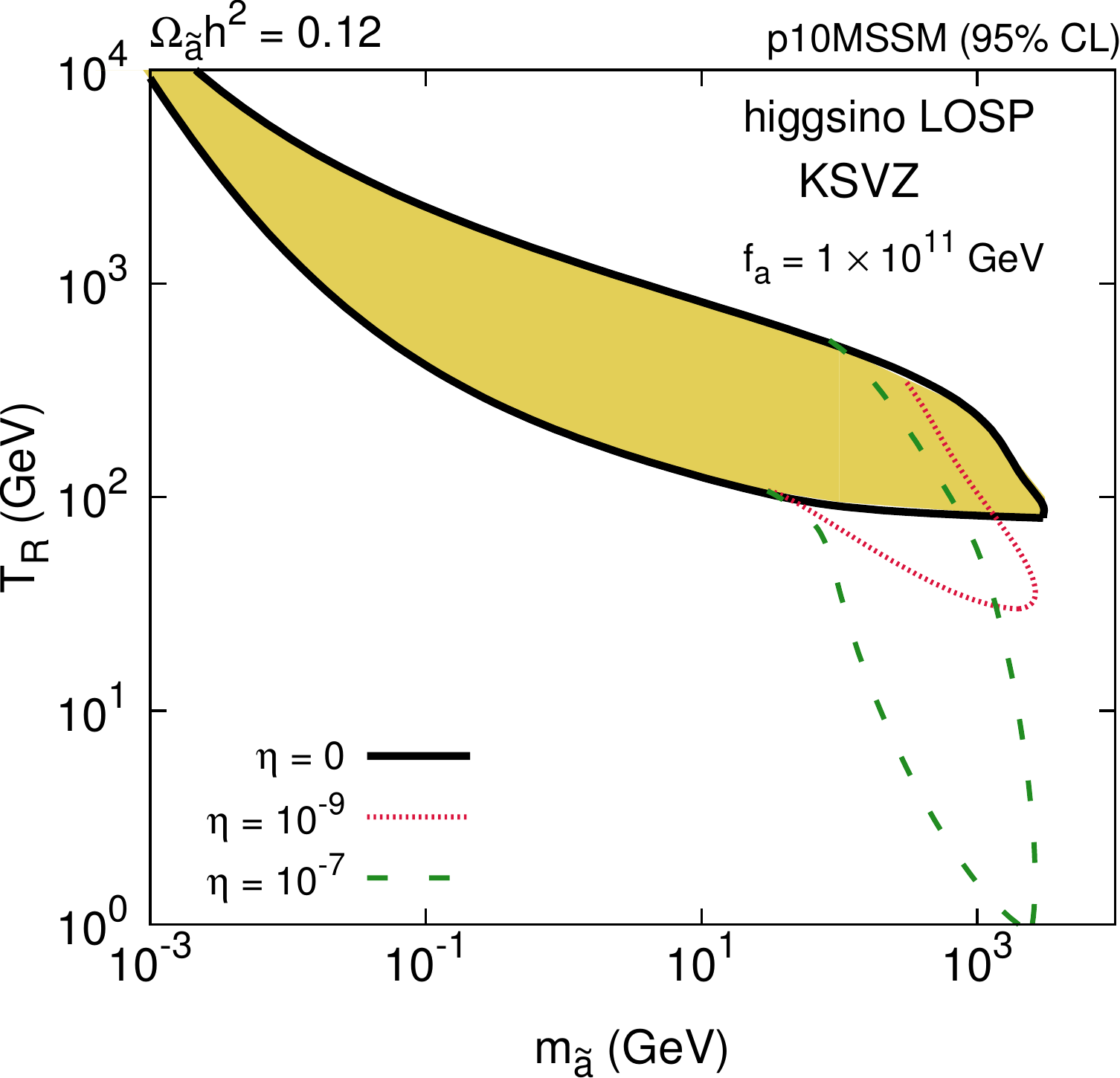}
\end{center}
\caption{%
The impact of direct and cascade decays of the inflaton on the allowed regions of $(m_{\tilde a},T_R)$ plane for
bino LOSP (left panel) and higgsino LOSP (right panel) in terms of the dimensionless parameter $\eta=b\cdot(100\,\mathrm{TeV}/m_\phi)$
defined in the text.
\label{fetanonzero}
}
\end{figure}

Our results are presented in Fig.\ \ref{fstau} and \ref{fetanonzero} where we plot the allowed regions for the stau, bino and higgsino LOSP for
different values of the dimensionless parameter $\eta=b\cdot(100\,\mathrm{TeV}/m_\phi)$,
where $b$ is an average number of axino DM particles per inflaton decay and $m_\phi$ is the inflaton mass at the minimum
of the potential. As inflaton decays provide an additional non-thermal component of axino DM (see a recent discussion in the case of gravitino \cite{Roszkowski:2014lga}), the allowed region becomes extended 
towards smaller values of $T_R$ at largest allowed values of $m_{\tilde a}$. This is because the additional NTP from inflaton decays allows for
a smaller contribution to axino DM density originating from LOSP decays, hence -- for a fixed set of the MSSM parameters -- for a smaller
$T_R$ and a larger suppression of LOSP abundance by entropy production in inflaton decays.

\section{Conclusions and outlook}
\label{sec:disc}

In this paper we have studied the impact of a low reheating temperature $T_R$ on thermal and non-thermal production of axino DM,
taking into account the non-instantaneous nature of the reheating process. We also extended previous studies by analyzing wide
ranges of phenomenologically acceptable parameters of the 10-parameter version of phenomenological MSSM 
instead of presenting the results for a single typical
parameter choice. Comparing our results with previous works, we found a number of differences in the allowed ranges 
of the axino mass $m_{\tilde a}$ and the reheating temperature. In particular, depending on the choice of the axion model
and the choice of the MSSM parameters, we showed that BBN constraints can exclude large portions of the parameter space
corresponding mainly to non-thermal production of axino DM relevant for low $T_R$. 
We also demonstrated how entropy production during reheating affects the upper limits on the axino mass for a given range of the MSSM parameters.

The are a few directions in which the analysis presented herein could be extended. In our work, we relaxed the simplified assumption of
instantaneous reheating, we still required that the maximum temperature during the inflaton-dominated period was sufficiently
larger than the reheating temperature that the supersymmetric particles could reach thermal equilibrium.  Although it is
a realistic requirement, one can also envision scenarios with a very small energy density at the end of inflation. 
Additionally, it has recently been noted that the maximum temperature during reheating 
may not be as large as previously estimated \cite{Mukaida:2015ria}. Such cases are not included in our study and we leave them for
future work.

An issue which requires some care in models with low $T_R$ is the origin of the primordial baryon asymmetry.
Since we work in a supersymmetric setup, the Affleck-Dine mechanism
\cite{Affleck:1984fy} (see, e.g., \cite{Allahverdi:2012ju} for a review)
is a feasible options, though a detailed discussion is beyond the scope of our study.

%%%%%%%%%%%%%%%%%%%%%%%%%%%%%%%%%%%%%%%%%%%%%%%%%

\paragraph*{Acknowledgements.}
We thank K.-Y.~Choi, L.~Covi and K.~Kohri for discussions and comments. ST would also like to thank E.~M.~Sessolo for a helpfull discussion regarding the LHC constraints.
This work has been funded in part by the Welcome Programme of the Foundation for Polish Science. 
LR is also supported in part by a STFC consortium grant of Lancaster, Manchester, and Sheffield Universities. ST is also partly supported by the National Science Centre, Poland, under research grant DEC-2014/13/N/ST2/02555. The use of the CIS computer cluster at the National Centre for Nuclear Research is gratefully acknowledged.
% KT is partially supported by the National Science Centre (Poland) grant 2014/14/E/ST9/00152. 

%%%%%%%%%%%%%%%%%%%%%%%%%%%%%%%%%%%%%%%%%%%%%%%%%

\section*{Appendix A: Axion and axino interactions}

The effective axion interaction Lagrangian after integrating out all heavy PQ charged fields can be written, to the lowest order terms in $1/f_a$, as
\begin{eqnarray}
\mathcal{L}_{a,\textrm{int}}^{\textrm{eff}} & = & c_1\,\frac{(\partial_\mu a)}{f_a}\,\Sigma_q{\bar{q}\,\gamma^\mu\,\gamma_5\,q} - \Sigma_q{(\bar{q}_L\,m\,q_R\,e^{i\,c_2\,a/f_a}+h.c.)}\nonumber\\
 & & + \frac{c_3}{32\pi^2\,f_a}\,a\,G\,\tilde{G} + \frac{C_{aWW}}{32\pi^2\,f_a}\,a\,W\,\tilde{W} + \frac{C_{aYY}}{32\pi^2\,f_a}\,a\,B\,\tilde{B} + \mathcal{L}_{\textrm{leptons}},
\label{Lagreffaxion}
\end{eqnarray}
where (following a partial integration over on-shell quark fields) the $c_1$ term can be reabsorbed into the $c_2$ term. The KSVZ case can be identified with $c_1 = 0$, $c_2 = 0$, $c_3\neq 0$, while the DFSZ one with $c_1 = 0$, $c_3 = 0$, $c_2\neq 0$. General axion models can have both $c_2\neq 0$ and $c_3\neq 0$.
$C_{aWW}$ and $C_{aYY}$ are model-dependent parameters that correspond to axino-gaugino-gauge boson anomaly interactions for the $U(1)_Y$ and the $SU(2)_L$ groups, respectively.

In a supersymmetrized version of an axion model~\cite{Nilles:1981py,Tamvakis:1982mw,Frere:1982sg} the real scalar axion field $a$ resides in a chiral supermultiplet since it is a gauge singlet. The other members of the axion supermultiplet are the fermionic superpartner axino $\tilde{a}$ and the real scalar field \textsl{saxion} $s$ that provides a remaining bosonic degree of freedom on-shell.

The interaction Lagrangian for the axion supermultiplet can be obtained by supersymmetrizing Eq.~(\ref{Lagreffaxion}). In particular, the axino-gaugino-gauge boson and the axino-gaugino-sfermion-sfermion interaction terms are given by~\cite{Covi:2001nw,Choi:2011yf}
\begin{eqnarray}
\mathcal{L}^{\textrm{eff}}_{\tilde{a}} & = & i\frac{\alpha_s}{16\pi\,f_a}\,\bar{\tilde{a}}\,\gamma_5\,[\gamma^\mu,\gamma^\nu]\,\widetilde{g}^b\,G^b_{\mu\nu} + \frac{\alpha_s}{4\pi\,f_a}\,\bar{\tilde{a}}\,\tilde{g}^a\,\Sigma_{\tilde{q}}{g_s\,\tilde{q}^\ast\,T^a\,\tilde{q}}\nonumber\\
 & & + i\frac{\alpha_2\,C_{aWW}}{16\pi\,f_a}\,\bar{\tilde{a}}\,\gamma_5\,[\gamma^\mu,\gamma^\nu]\,\widetilde{W}^b\,W^b_{\mu\nu} + \frac{\alpha_2}{4\pi\,f_a}\,\bar{\tilde{a}}\,\widetilde{W}^a\,\Sigma_{\tilde{f}_D}{g_2\,\tilde{f}_D^\ast\,T^a\,\tilde{f}_D}\nonumber\\
 & & + i\frac{\alpha_Y\,C_{aYY}}{16\pi\,f_a}\,\bar{\tilde{a}}\,\gamma_5\,[\gamma^\mu,\gamma^\nu]\,\widetilde{B}\,B_{\mu\nu} + \frac{\alpha_Y}{4\pi\,f_a}\,\bar{\tilde{a}}\,\widetilde{B}\,\Sigma_{\tilde{f}}{g_Y\,\tilde{f}^\ast\,Q_Y\,\tilde{f}},
\label{eqaxinoLint}
\end{eqnarray}
where $\tilde{f}_D$ and $\tilde{f}$ denote sfermions carrying non-zero $T^3$ and $Y$, respectively.  

A generic form of interactions between the axion and matter supermultiplets was considered in~\cite{Bae:2011jb}. In particular, it was pointed out that, for $v_{\textrm{PQ}} > T\gtrsim \,M_\Phi$, where $M_\Phi$ is the mass of the heaviest PQ-charged and gauge-charged supermultiplet $\Phi$, the axino-gaugino-gauge boson interaction term is suppressed by $M_{\Phi}^2/T^2$. This is particularly important for the DFSZ axino, where $\Phi$ corresponds to the Higgs supermultiplets and therefore $M_\Phi = \mu$ (the higgsino mass). The dominant contribution to axino TP is then associated with a higgsino decay to the axino and the Higgs boson that is described by~\cite{Bae:2011jb,Kim:1983dt,Chun:2011zd}
\begin{equation}
\mathcal{L}^{\textrm{eff}}_{\tilde{a},\textrm{DFSZ}}\ni c_H\,\frac{\mu}{f_a}\,\tilde{a}\,[\widetilde{H}_d\,H_u + \widetilde{H}_u\,H_d] + h.c.
\label{eqaxinoDFSZLint}
\end{equation}

\section*{Appendix B: Calculation of axino TP with low reheating temperature}
\label{sec:appe}

In scenarios with non-instantaneous reheating in calculating $Y_{\tilde{a}}^{\textrm{TP}}$ one has to take into account a modified expansion rate of the Universe. Below we briefly describe the methodology that can be used to calculate the axino TP yield.

\paragraph*{Axino TP yield with non-instantaneous reheating.}  
It results in a modification of temperature dependence on the scale factor $T(a)$. The Boltzmann equation can then be written as
\begin{equation}
\frac{\mathrm{d}X_{\tilde{a}}}{\mathrm{d}T}\,\frac{\mathrm{d}T}{\mathrm{d}a} = \frac{a^2}{H}\Big(\Sigma_{\textrm{scat}} + \Sigma_{\textrm{dec}}\Big),
\end{equation}
where $X_{\tilde{a}} = a^3\,n_{\tilde{a. }}$. In that case the present-day axino abundance can be written as:
\begin{equation}
Y_{\tilde{a},0} =\frac{1}{s_0A_0^3}\int_{T_0}^{T_{\textrm{up}}}{\mathrm{d}T\,\Big(-T\frac{\mathrm{d}\ln{T}}{\mathrm{d}A}\Big)^{-1}\,\frac{A^2}{H}\Big(\Sigma_{\textrm{scat}} + \Sigma_{\textrm{dec}}\Big)},
\label{Yrehper}
\end{equation}
where $T_{\textrm{up}}$ corresponds to an effective upper limit in the integration (in practice it is sufficient to use $T_{\textrm{up}} \simeq (5- 10)T_{\textrm{RD}}$, as for larger temperatures TP of axinos is more efficient, but the fast expansion of the Universe in the reheating period dilutes away all the axinos produced at that early times). We also used $A = a/a_I = aT_R$  and  
\begin{equation}
\frac{\mathrm{d}\,\ln T}{\mathrm{d}A} = \frac{\frac{1}{4R} \frac{\mathrm{d}R}{\mathrm{d}A}-\frac{1}{A}}{1+\frac{1}{4} \frac{\mathrm{d}\,\ln g_\ast(T)}{\mathrm{d}\,\ln T}} \, ,
\end{equation}
where $R$ is related to the energy density of radiation $\rho_R$ by $R=\rho_R a^4$.

\paragraph*{The scattering term.} In order to deal with the scattering contribution to $Y_{\tilde{a}}^{\textrm{TP}}$ we 
express $\Sigma_{\textrm{scat}}$ in terms of the scattering cross-section $\sigma(s)$, following \cite{Choi:1999xm}), and obtain:
\begin{eqnarray}
\hspace{-2cm}Y_{\tilde{a},0}^{\textrm{scat},i,j} &=& \frac{1}{s_0A_0^3}\,\frac{g_ig_j}{16\pi^4}\int_{T_0}^{T_{\textrm{up}}} \mathrm{d}T \int_{(m_1+m_2)/T}^{\infty}\mathrm{d}x\,\Bigg[\Big(-T\frac{\mathrm{d}\ln{T}}{\mathrm{d}A}\Big)^{-1}\,\frac{A^2}{H}\Bigg]
\times\nonumber\\
& & \hspace{0.6cm}
T^2\,K_1(x)\,\sigma(x^2T^2)\Big[\big(x^2T^2-m_1^2-m_2^2\big)^2-4m_1^2m_2^2\Big].
\label{eq:appe1}
\end{eqnarray}
We then change the order of integration and decompose the above integral into
\begin{equation}
Y_{\tilde{a},0}^{\textrm{scat},i,j} = \frac{1}{s_0A_0^3}\,\frac{g_ig_j}{16\pi^4}\,\Big(I_1+I_2\Big),
\end{equation}
where $I_1$ and $I_2$ are given by expressions very similar to (\ref{eq:appe1}), but with $T$
integrated from $T_0$ to $(m_1+m_2)/x$ and from $(m_1+m_2)/x$ to $T_{\textrm{up}}$, respectively.
Then $I_2\approx 0$
since $T_0\approx 0$. For the remaining integral $I_1$ one obtains
\begin{eqnarray}
\hspace{-1.5cm}Y_{\tilde{a},0}^{\textrm{scat},i,j}& \simeq & \frac{\bar{g}\,g_i\,g_j\,M_{\textrm{Pl}}}{16\pi^4}\,\int^\infty_{(m_1+m_2)/T_{\textrm{up}}}\mathrm{d}t\,t^3K_1(t)
\int_{m_1+m_2}^{tT_{\textrm{up}}}\mathrm{d}(\sqrt{s})\,\nonumber\\
 & & \hspace{2.7cm}f(\sqrt{s})\,\sigma(s)\,\frac{\big(s-m_1^2-m_2^2\big)^2-4m_1^2m_2^2}{s^2},
\label{gen_scat}
\end{eqnarray}
where $\bar{g} = \frac{135\sqrt{10}}{2\pi^3\,g_\ast^{3/2}}$ and
\begin{equation}
f(\sqrt{s}) = \frac{\pi}{T_0^3A_0^3}\sqrt{\frac{g_\ast}{30}}\,\Big(-T\frac{\mathrm{d}\ln{T}}{\mathrm{d}A}\Big)^{-1}\,\frac{A^2T^6}{\sqrt{\frac{\Phi\,T_R^4}{A^3} + \frac{RT_R^4}{A^4}}},\hspace{1cm}\textrm{with }T=\frac{\sqrt{s}}{t}.
\label{eqfunfrehperiod}
\end{equation}
A careful analysis of Eq.~(\ref{eqfunfrehperiod}) in the reheating period shows that 
one can make an approximation
\begin{equation}
f = \left\{\begin{array}{ll}
\Big({T_{\textrm{RD}}}/{T}\Big)^{-c} & (\leq 1)\textrm{  in the reheating period},\\
1 & \textrm{in the RD epoch},
\end{array}\right.
\label{final2f}
\end{equation}
where $c\simeq -7$ and $T_{\textrm{RD}}\sim 0.5\,T_R$. In practice we find more exact values of $a$ and $T_{\textrm{RD}}$ numerically, but they depend only slightly on the model parameters.

\paragraph*{High ${T_R}$ limit of the scattering term.} The integral (\ref{gen_scat}) can be rewritten as a sum of three integrals,
schematically represented by
\begin{equation}
Y_{\tilde{a},0}^{\textrm{scat},i,j} \sim \int_{(m_1+m_2)/T_{\textrm{up}}}^{(m_1+m_2)/T_{\textrm{RD}}}
\int_{m_1+m_2}^{tT_{\textrm{up}}} + \int_{(m_1+m_2)/T_{\textrm{RD}}}^{\infty}
\int_{m_1+m_2}^{tT_{\textrm{RD}}} + \int_{(m_1+m_2)/T_{\textrm{RD}}}^{\infty}
\int_{tT_{\textrm{RD}}}^{tT_{\textrm{up}}} = J_1 + J_2 + J_3.
\end{equation}
One can verify that in the limit $T_{\textrm{RD}}\rightarrow\infty$, we have $J_1\rightarrow0$, since the range of  external integration shrinks to zero , while the integrand does not diverge. The second integral, $J_2$, corresponds to the the standard result obtained in the instantaneous reheating approximation. In the limit of high reheating temperature the inner integral can be simplified to
\begin{eqnarray}
\int^{tT_{\textrm{RD}}}_{m_1+m_2}{\mathrm{d}(\sqrt{s})\,f(s)\,\sigma(s,t)\,\frac{(s-m_1^2-m_2^2)^2-4m_1^2m_2^2}{s^2}} &\simeq& \int^{tT_{\textrm{RD}}}_{m_1+m_2}{\mathrm{d}(\sqrt{s})\,1\times\sigma(t)\times 1}\nonumber\\
& \simeq & t\,\sigma(t)\,T_{\textrm{RD}},
\end{eqnarray}
($\sigma$ depends on $t$ via $m_{\textrm{eff}}$, see, \textsl{e.g.}, \cite{Choi:2011yf})
where we noticed that the integral is mainly determined by the values of the integrand in high-$s$ limit in which, to a good approximation, $\sigma(s,t) = \sigma(t)$.
For the third integral, $J_3$, we similarly note that inner integration leads to
\begin{equation}
\int_{tT_{\textrm{RD}}}^{tT_{\textrm{up}}}{\mathrm{d}(\sqrt{s})\,f(s)\,\sigma(s,t)\,\frac{(s-m_1^2-m_2^2)^2-4m_1^2m_2^2}{s^2}} \simeq \sigma(t)\,\int_{tT_{\textrm{RD}}}^{tT_{\textrm{up}}}{\mathrm{d}(\sqrt{s})\,f(s)}.
\label{eq:e18}
\end{equation}
In the integration range $T = \sqrt{s}/t>T_{\textrm{RD}}$ and therefore (\ref{eq:e18}) becomes
\begin{equation}
 t\,\sigma(t)\,\int_{T_{\textrm{RD}}}^{T_{\textrm{up}}}{\mathrm{d}T\,\Big(\frac{T_{\textrm{RD}}}{T}\Big)^7}\simeq \frac{1}{6}\,t\,\sigma(t)\,T_{\textrm{RD}},
\end{equation}
where we assumed $T_{\textrm{up}} = cT_{\textrm{RD}}$ with $c$ high enough so that effectively $T_{\textrm{up}}$ can be replaced by $\infty$ in the integration. The remaining (external) integrals for both $J_2$ and $J_3$ are the same. Hence for high $T_R$
\begin{equation}
\frac{J_3}{J_2}\simeq \frac{1}{6}\simeq 0.17
\label{eqhighTRCB}
\end{equation}
Eq.~(\ref{eqhighTRCB}) is valid for each contribution to the scattering term.

\paragraph*{The decay term.}
In the case of the decay term we substitute $\langle\Gamma\rangle\,n^{eq}_i$ (see, \textsl{e.g.}, \cite{Choi:1999xm}) into (\ref{Yrehper}) and obtain
\begin{equation}
Y_{\tilde{a},0}^{\textrm{dec},i} = \frac{1}{s_0A_0^3}\,\frac{\Gamma\,g_i\,m_i}{2\pi^2}\int_{T_0}^{T_{\textrm{up}}}{\mathrm{d}T\,\int_{m_i/T}^{\infty}{\mathrm{d}x\,\Bigg[\Big(-T\frac{\mathrm{d}\ln{T}}{\mathrm{d}A}\Big)^{-1}\,\frac{A^2}{H}\Bigg]\,T^2\,\frac{\sqrt{x^2-\frac{m_i^2}{T^2}}}{e^x\mp 1}}}.
\end{equation}
Once again we change the order of integration and find that one term is negligible, while the other leads to
\begin{equation}
Y_{\tilde{a},0}^{\textrm{dec},i} \simeq \frac{\bar{g}\,g_i\,\Gamma\,m_i\,M_{\textrm{Pl}}}{2\pi^2}\,\int_{m/T_{\textrm{up}}}^{\infty}{\mathrm{d}t\,\frac{t^4}{e^t\mp 1}\,\int_{m_i}^{tT_{\textrm{up}}}{\mathrm{d}E\,f(E)\,\frac{1}{E^4}\,\sqrt{1-\frac{m_i^2}{E^2}}}},
\end{equation}
where $f$ is given by Eq.~(\ref{final2f}) with $\sqrt{s}$ replaced by $E$. 
The inner integral
\begin{equation}
g_{m_i,T_R}(t) = \int_{m_i}^{tT_{\textrm{up}}}{\mathrm{d}E\,f(T=E/t)\,\frac{1}{E^4}\,\sqrt{1-\frac{m_i^2}{E^2}}},
\label{finalg}
\end{equation}
where $m_i/T_{\textrm{up}}\leq t\leq \infty$ can be calculated analytically. Depending on the value of $t$ one obtains
for $m_i/T_{\textrm{up}}\leq t\leq m_i/T_{\textrm{RD}}$
\begin{equation}
g_{m_i,T_R}(t) = g^{\textrm{reh}}_{m_i,T_R}(t) = \frac{T_{\textrm{RD}}^7\,t^7}{m_i^{10}}\,
\Big(\frac{1}{3}w^3-\frac{4}{5}w^5+\frac{6}{7}w^7-\frac{4}{9}w^9+\frac{1}{11}w^{11}\Big)\Bigg|_{0}^{\sqrt{1-\big[m_i/(tT_{\textrm{up}})\big]^2}}.
\label{gsol1}
\end{equation}
and for $t\geq m_i/T_{\textrm{RD}}$ (the temperature can be either larger or smaller than $T_{\textrm{RD}}$) 
\begin{equation}
g_{m_i,T_R}(t) = g^{\textrm{RD}}_{m_i,T_R}(t) + g^{\textrm{reh}}_{m_i,T_R}(t),
\label{gsol2}
\end{equation}
where ($t_R = m_i/T_{\textrm{RD}}$)
\begin{equation}
g^{\textrm{RD}}_{m_i,T_R}(t) = \frac{1}{8m_i^3}\,\Big(\frac{\pi}{2}-\arctan{\frac{t_R}{\sqrt{t^2-t_R^2}}}+\frac{t_R}{t^4}(t^2-2t_R^2)\sqrt{t^2-t_R^2}\Big),
\label{gsol2a}
\end{equation}
\begin{equation}
g^{\textrm{reh}}_{m_i,T_R}(t) = \frac{T_{\textrm{RD}}^7\,t^7}{m_i^{10}}\,
\Big(\frac{1}{3}w^3-\frac{4}{5}w^5+\frac{6}{7}w^7-\frac{4}{9}w^9+\frac{1}{11}w^{11}\Big)\Bigg|_{\sqrt{1-\big[m_i/(tT_{\textrm{RD}})\big]^2}}
^{\sqrt{1-\big[m_i/(tT_{\textrm{up}})\big]^2}},
\label{gsol2b}
\end{equation}
One can verify that in the case of instantaneous reheating the standard result~\cite{Choi:1999xm} is rederived.

% ----------------------------------------

\section*{Appendix C: Phase space integrals for 
%free-streaming length of 
non-thermally produced axinos}
\label{sec:numerical}

\begin{figure}[ht]
\begin{center}
\includegraphics*[width=6cm]{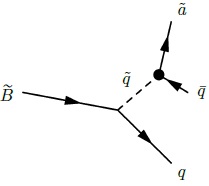}
\hspace{0.5cm}
\includegraphics*[width=6cm]{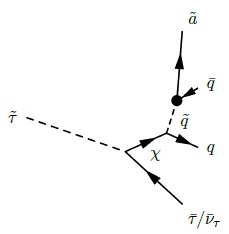}
\end{center}
\caption{Feynman diagrams for the bino LOSP (left panel) and the stau LOSP (right panel) decaying into the axino for $C_{aYY}=0$.}
\label{Fig:feyn}
\end{figure}

In this appendix we provide results for both the LOSP lifetime and the present-day rms velocity of axinos in the case of $C_{aYY}=0$. Depending on the nature of the LOSP this requires analysis of $3-$ of $4-$body decays.

\textbf{$\mathbf{3-}$body bino decay to the axino with} $\mathbf{C_{aYY}=0}$\ \ We calculate the lifetime that corresponds to the $3-$body decay shown in the left panel of Fig.~\ref{Fig:feyn}. In the case of $m_{\tilde{a}}\ll m_{\widetilde{B}}$ one obtains
\begin{equation}
\tau \simeq 30\,\textrm{sec}\,\frac{\left(\frac{100\,\textrm{GeV}}{m_{\widetilde{B}}}\right)^5\,\left(\frac{m_{\tilde{q}}}{1\,\textrm{TeV}}\right)^4\,\left(\frac{1\,\textrm{TeV}}{m_{\tilde{g}}}\right)^2\,\left(\frac{f_a}{10^{11}\,\textrm{GeV}}\right)^2}{y^2\left\{-\frac{5}{2} + 3y + (3y^2-4y+1)\ln\left[{1-\frac{1}{y}}\right]\right\}},\hspace{1cm}\textrm{where\ \ }y = m^2_{\tilde{q}}/m^2_{\widetilde{B}}.
\label{eq:cayy0nonzeromass}
\end{equation}
It is straightforward to verify that Eq.~(\ref{eq:cayy0nonzeromass}) can be simplified to Eq.~(\ref{eq:cayy0}) for $m^2_{\widetilde{B}}\ll m_{\tilde{q}}^2$.

When treating the WDM constraint we calculate the present-day rms velocity\ \cite{Jedamzik:2005sx} by
\begin{equation}
\langle v_{\tilde{a}}^0\rangle = 4.57\times 10^{-5}\,\frac{\textrm{km}}{\textrm{s}}\,g_d^{-1/12}\,\frac{\langle |p_{\tilde{a}}|\rangle}{m_{\tilde{a}}}\,\left(\frac{\tau_{\widetilde{B}}}{1\textrm{s}}\right)^{1/2},
\end{equation}
where $\langle |p_{\tilde{a}}\rangle|$ is the average momentum of the outgoing axino. The final result reads
\begin{equation}
\langle v_{\tilde{a}}^0\rangle = \left(2.5\times 10^{-3}\,\frac{\textrm{km}}{\textrm{s}}\right)\,g_d^{-1/12}\,\left(\frac{1\,\textrm{GeV}}{m_{\tilde{a}}}\right)\,\sqrt{\frac{m_{\widetilde{B}}}{100\,\textrm{GeV}}}\,\left(\frac{1\,\textrm{TeV}}{m_{\tilde{g}}}\right)\,\left(\frac{f_a}{10^{11}\,\textrm{GeV}}\right)\times f\left(\frac{m_{\tilde{q}}^2}{m_{\widetilde{B}}^2}\right),
\end{equation}
where
\begin{equation}
f(y) = \frac{\langle |p_{\tilde{a}}|\rangle}{m_{\widetilde{B}}}\,\frac{1}{\sqrt{-\frac{5}{2} + 3y + (3y^2-4y+1)\ln\left[{1-\frac{1}{y}}\right]}} \simeq 1.22\,y-0.47.
\end{equation}
The numerical approximation in the last equation works well unless $m_{\widetilde{B}}\simeq m_{\tilde{q}}$ for which function $f$ becomes suppressed.
\medskip

\begin{figure}[ht]
\begin{center}
\includegraphics*[width=8cm,height=7cm]{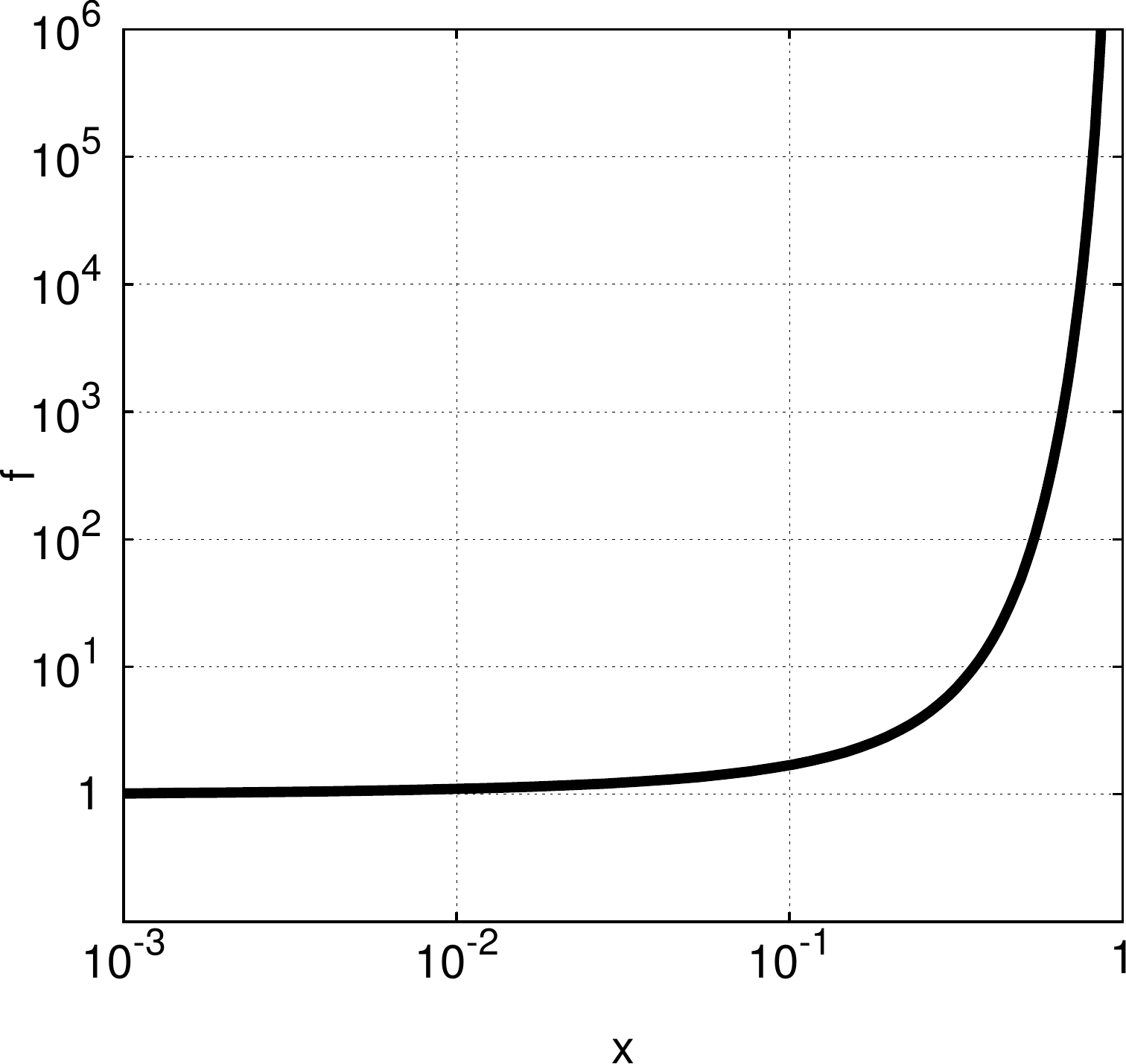}
\hspace{0.5cm}
\includegraphics*[width=8cm,height=7cm]{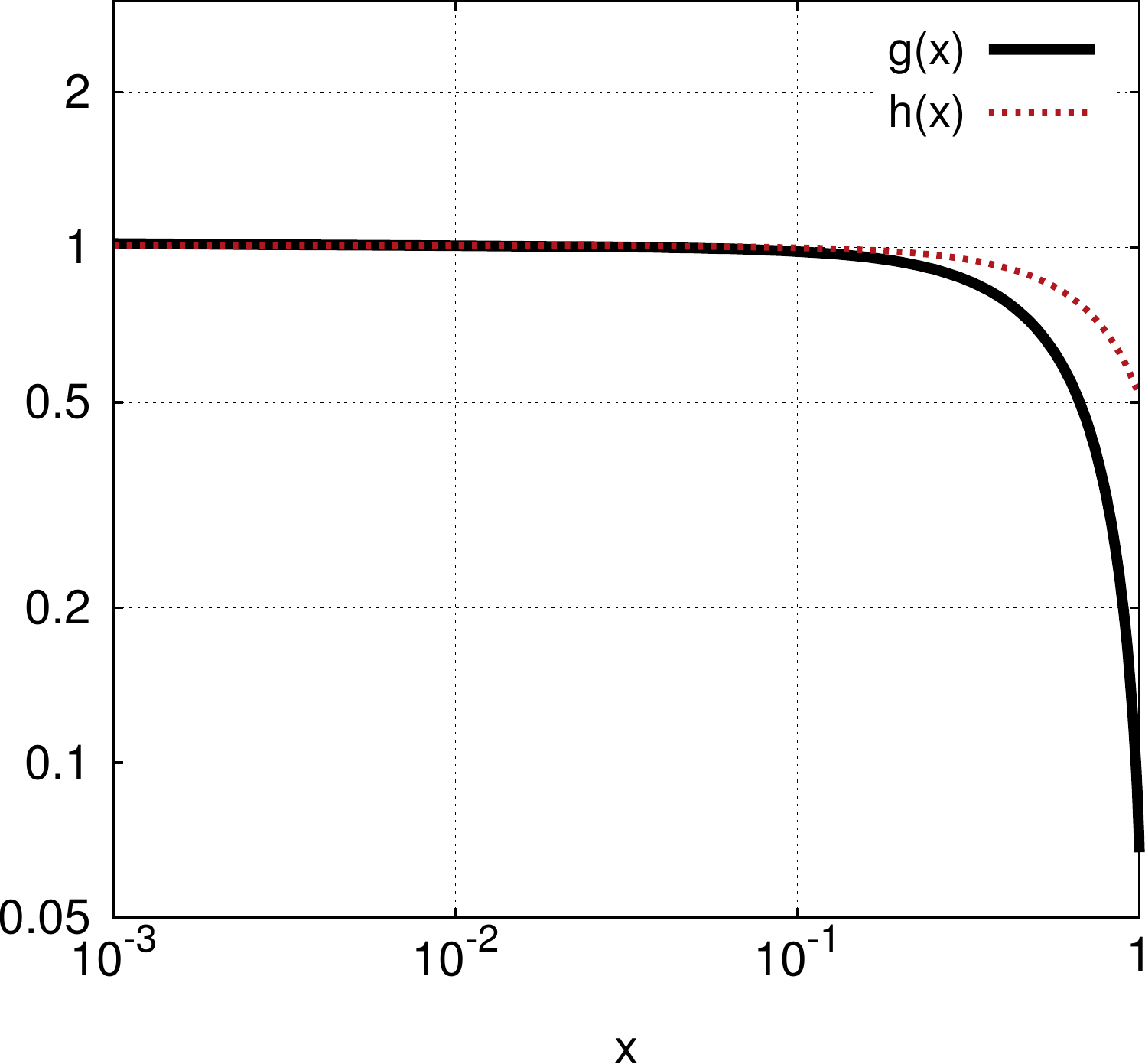}
\end{center}
\caption{Auxiliary functions in the formula for the lifetime of stau decaying into axino if $C_{aYY}=0$.}
\label{Fig:fgh}
\end{figure}

\textbf{$\mathbf{4-}$body stau decay to the axino with} $\mathbf{C_{aYY}=0}$\ \ The respective Feynman diagram is shown in the right panel of Fig.~\ref{Fig:feyn}. The final result for the lifetime of the stau reads
\begin{eqnarray}
\tau &\simeq & C_{L/R}\,(0.051\,\textrm{sec})\times\left(\frac{100\,\textrm{GeV}}{m_{\tilde{\tau}}}\right)^{10}\,\left(\frac{m_{\widetilde{B}}}{100\,\textrm{GeV}}\right)^{4}\,\left(\frac{m_{\tilde{q}}}{1\,\textrm{TeV}}\right)^{4}\,\left(\frac{1\,\textrm{TeV}}{m_{\tilde{g}}}\right)^{2}\,\left(\frac{f_a}{10^{11}\,\textrm{GeV}}\right)^{2}\times\nonumber\\
& & \hspace{8cm}\times f\left(\frac{m_{\tilde{a}}}{m_{\tilde{\tau}}}\right)\,g\left(\frac{m_{\tilde{\tau}}}{m_{\widetilde{B}}}\right)\,h\left(\frac{m_{\tilde{\tau}}}{m_{\tilde{q}}}\right),
\end{eqnarray}
where $C_R = 1$ for the ''right'' stau and $C_L\simeq 4.45$ for the ''left'' stau, while functions $f(x)$, $g(x)$ and $h(x)$ are shown in Fig.~\ref{Fig:fgh}. They all tend to $1$ for $x\ll 1$, i.e., for $m_{\tilde{a}}\ll m_{\tilde{\tau}}\ll m_{\widetilde{B}},m_{\tilde{q}}$. 

\begin{figure}[!ht]
\begin{center}
\includegraphics*[width=8cm,height=7cm]{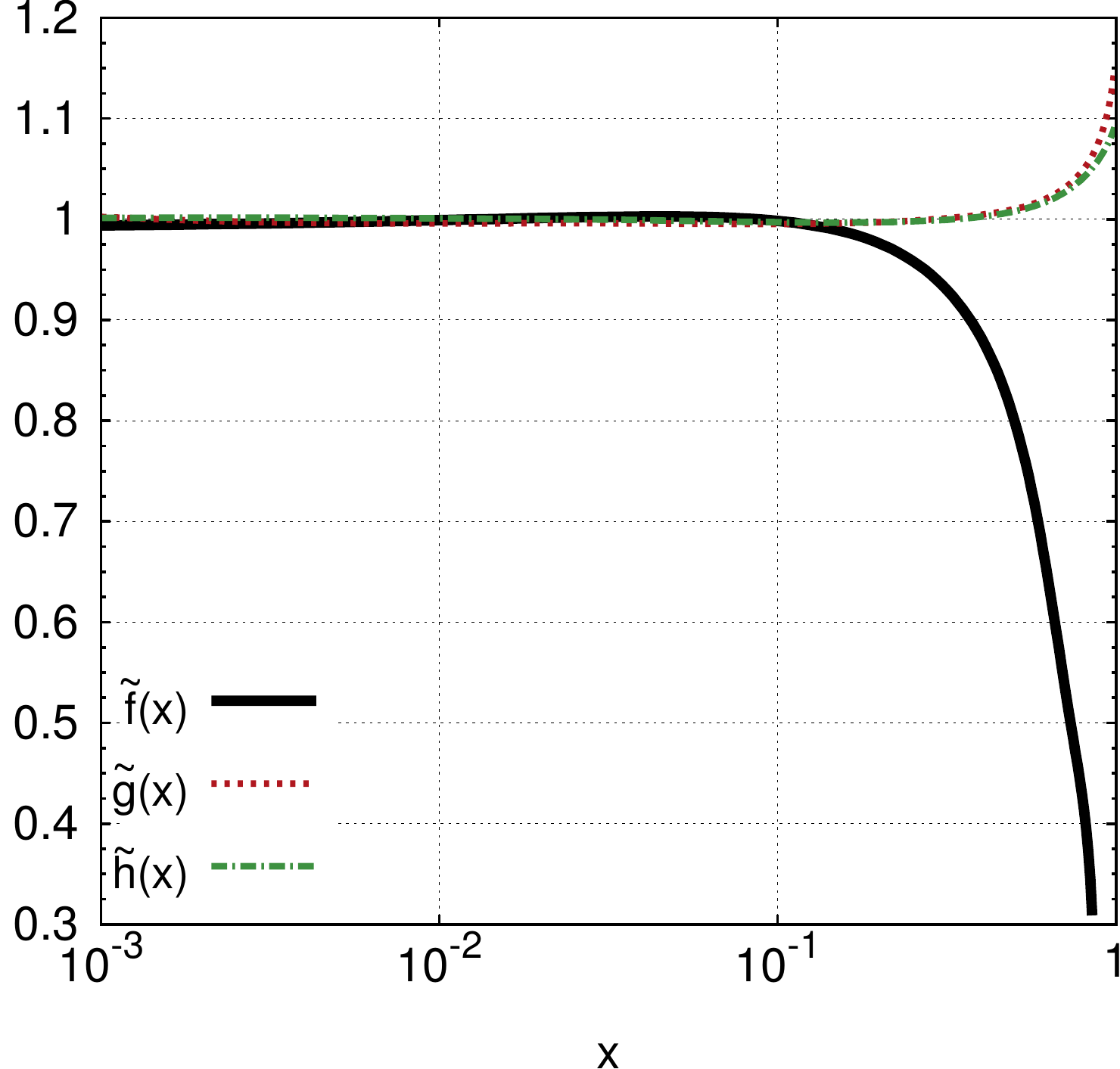}
%\hspace{0.5cm}
%\includegraphics*[width=8cm,height=7cm]{gh.eps}
\end{center}
\caption{Auxiliary functions in the formula for the present-day rms velocity of axinos produced in late-time stau decays if $C_{aYY}=0$.}
\label{Fig:tilfgh}
\end{figure}

The present-day rms velocity is equal to
\begin{eqnarray}
v_{\tilde{a}}^0 & \simeq & \left(2.58\times 10^{-4}\,\frac{\textrm{km}}{\textrm{s}}\right)g_d^{-1/12}\,\,\widetilde{C}_{L/R}\times  \tilde{f}\left(\frac{m_{\tilde{a}}}{m_{\tilde{\tau}}}\right)\,\tilde{g}\left(\frac{m_{\tilde{\tau}}}{m_{\widetilde{B}}}\right)\,\tilde{h}\left(\frac{m_{\tilde{\tau}}}{m_{\tilde{q}}}\right)\times\nonumber\\
& & \times\left(\frac{1\,\textrm{GeV}}{m_{\tilde{a}}}\right)\left(\frac{100\,\textrm{GeV}}{m_{\tilde{\tau}}}\right)^{4}\,\left(\frac{m_{\widetilde{B}}}{100\,\textrm{GeV}}\right)^{2}\,\left(\frac{m_{\tilde{q}}}{1\,\textrm{TeV}}\right)^{2}\,\left(\frac{1\,\textrm{TeV}}{m_{\tilde{g}}}\right)\,\left(\frac{f_a}{10^{11}\,\textrm{GeV}}\right),
\end{eqnarray}
where $\widetilde{C}_R = 1$ for the ''right'' stau and $\widetilde{C}_L = 2.3$ for the ''left'' stau while functions $\tilde{f}$, $\tilde{g}$ and $\tilde{h}$ are shown in Fig.~\ref{Fig:tilfgh}.

% ---------------------------------------

\section*{Appendix D: Description of the numerical analysis}
\label{sec:num}

%%%%%%%%%%%%%%%%%%
\begin{table}[ht]
\centering
\begin{tabular}{|c|c|}
\hline 
Parameter & Range \\ 
\hline
\hline 
bino mass & $0.1 < M_1 < 5$ \\ 
wino mass & $0.1 < M_2 < 6$ \\ 
gluino mass & $0.7 < M_3 < 10$ \\ 
stop trilinear coupling & $-12 < A_t < 12$ \\ 
stau trilinear coupling & $-12 < A_{\tau} < 12$ \\ 
sbottom trilinear coupling & $A_b = -0.5$ \\ 
pseudoscalar mass & $0.2 < m_A < 10$ \\ 
$\mu$ parameter & $0.1 < \mu < 6$ \\ 
3rd gen. soft squark mass & $0.1 < m_{\widetilde{Q}_3} < 15$ \\ 
3rd gen. soft slepton mass & $0.1 < m_{\widetilde{L}_3} < 15$ \\ 
1st/2nd gen. soft squark mass  & $m_{\widetilde{Q}_{1,2}} = m_{\widetilde{Q}_3} + 1$ TeV \\ 
1st/2nd gen. soft slepton mass  & $m_{\widetilde{L}_{1,2}} = m_{\widetilde{L}_3} + 100$ GeV \\ 
ratio of Higgs doublet VEVs & $2 < \tan\beta < 62$ \\ 
\hline 
\hline
Nuisance parameter & Central value, error \\
\hline
\hline
Bottom mass \mbmbmsbar (GeV) & (4.18, 0.03) \cite{PDG:2014}\\
Top pole mass \mtop (GeV) & (173.5, 1.0) \cite{PDG:2014}\\
\hline
\end{tabular}
\caption{The parameters of the p10MSSM and their
  ranges used in our scan. All masses and trilinear couplings are given in TeV, unless
  indicated otherwise. All the parameters of the model are given at the
  SUSY breaking scale.} 
\label{tabp10MSSM} 
\end{table}
%%%%%%%%%%%%%%%%%%

%%%%%%%%%%%%%%%%%%
\begin{table}[ht]
\centering
\begin{tabular}{|c|c|c|c|}
\hline 
Measurement & Mean & Error: exp., theor. & Ref. \\ 
\hline
\hline 
$m_h$ & $125.7$ GeV  & $0.4$ GeV, $3$ GeV & \cite{CMS:2013}\\ 
\abunchi & $0.1199$ & $0.0027$, $10\%$ & \cite{Planck:2013}\\ 
\brbxsgamma$\times 10^4$ & $3.43$ & $0.22$, $0.21$ & \cite{SLAC} \\ 
\brbutaunu$\times 10^4$ & $0.72$ & $0.27$, $0.38$ & \cite{Belle:2012}\\ 
\delmbs & $17.719$ $\textrm{ps}^{-1}$ & $0.043\,\textrm{ps}^{-1}$, $2.400\,\textrm{ps}^{-1}$ & \cite{PDG:2014}\\
\sinsqeff & $0.23116$& $0.00013$, $0.00015$ & \cite{PDG:2014}\\
$M_W$ & $80.385$ GeV & $0.015$ GeV, $0.015$ GeV & \cite{PDG:2014}\\
\brbsmumu$\times 10^9$ & $2.9$ & $0.7$, $10\%$ & \cite{LHCb:2013,CMS:2013Jul}\\
\hline 
\end{tabular}
\caption{The constraints imposed on the parameter spaces of the p10MSSM and
the  CMSSM. The LUX upper limits \cite{LUX:2013} have been implemented as a hard cut.}
\label{constraints} 
\end{table}
%%%%%%%%%%%%%%%%%%

Here we explain some details of our numerical analysis of the scenario of axino DM with low reheating
temperatures of the Universe in the context of the MSSM.  
A study of a completely general MSSM would be at the same time
complicated and unnecessary, hence we
select a 10-parameter version of the MSSM
(p10MSSM) which has practically all the relevant features of the general model.  
These adjustable parameters of the model and
their ranges are specified in Table~\ref{tabp10MSSM}. Our choice is closely related to
that of \cite{Fowlie:2013} (see discussion therein), except that we
keep both the wino mass $M_2$ and the bino mass
$M_1$ free. 

We scan the parameter space of p10MSSM following the Bayesian
approach. The numerical analysis was performed using the BayesFITS package
which utilizes Multinest \cite{Feroz:2009} for sampling the parameter
space of the model. Mass spectra were calculated with
SOFTSUSY-3.4.0 \cite{Allanach:2002}, while $B$-physics related
quantities with SuperIso v3.3 \cite{Arbey:2009}. 

The constraints imposed  in scans are listed in
Table~\ref{constraints}. The LHC limits for supersymmetric particle
masses were implemented following the methodology described in
\cite{Fowlie:2013,Fowlie:2012}.  The DM relic density for low $T_R$
was calculated by solving numerically the set of Boltzmann equations, 
as outlined in \cite{Giudice:2000,Roszkowski:2014lga}. In order to
find the point where WIMPs freeze out, we adapted the method described,
e.g., in \cite{Dev:2013yza} to the scenario with a low reheating
temperature, extracting the relevant functions entering the Boltzmann equations ($\langle\sigma v\rangle_{\textrm{eff}}$ and
$\langle\sigma v\rangle_{\textrm{eff}}\langle E\rangle_{\textrm{eff}}$, cf.\ Ref.~\cite{Roszkowski:2014lga})
with appropriately modified
MicrOMEGAs v3.6.7 \cite{Belanger:2013}.

%%%%%%%%%%%%%%%%%%%%%%%%%%%%%%%%%%%%%%%%%%%%%%%%%


\begin{thebibliography}{99}

\bibitem{Baer:2014eja}
  H.~Baer, K.-Y.~Choi, J.~E.~Kim and L.~Roszkowski,
  %``Dark matter production in the early Universe: beyond the thermal WIMP paradigm,''
  Phys.\ Rept.\  {\bf 555} (2014) 1
  [arXiv:1407.0017 [hep-ph]].
  %%CITATION = ARXIV:1407.0017;%%
  
\bibitem{Peccei:1977hh}
  R.~D.~Peccei and H.~R.~Quinn,
  %``CP Conservation in the Presence of Instantons,''
  Phys.\ Rev.\ Lett.\  {\bf 38} (1977) 1440.
  %%CITATION = PRLTA,38,1440;%%

\bibitem{Peccei:1977ur}
  R.~D.~Peccei and H.~R.~Quinn,
  %``Constraints Imposed by CP Conservation in the Presence of Instantons,''
  Phys.\ Rev.\ D {\bf 16} (1977) 1791.
  %%CITATION = PHRVA,D16,1791;%%

\bibitem{Weinberg:1977ma}
  S.~Weinberg,
  %``A New Light Boson?,''
  Phys.\ Rev.\ Lett.\  {\bf 40} (1978) 223.
  %%CITATION = PRLTA,40,223;%%
  
\bibitem{Kim:2008hd}
  J.~E.~Kim and G.~Carosi,
  %``Axions and the Strong CP Problem,''
  Rev.\ Mod.\ Phys.\  {\bf 82} (2010) 557
  [arXiv:0807.3125 [hep-ph]].
  %%CITATION = ARXIV:0807.3125;%%

\bibitem{Bae:2008ue}
  K.~J.~Bae, J.~H.~Huh and J.~E.~Kim,
  %``Update of axion CDM energy,''
  JCAP {\bf 0809} (2008) 005
  [arXiv:0806.0497 [hep-ph]].
  %%CITATION = ARXIV:0806.0497;%%

\bibitem{Kim:1979if}
  J.~E.~Kim,
  %``Weak Interaction Singlet and Strong CP Invariance,''
  Phys.\ Rev.\ Lett.\  {\bf 43} (1979) 103.
  %%CITATION = PRLTA,43,103;%%

\bibitem{Shifman:1979if}
  M.~A.~Shifman, A.~I.~Vainshtein and V.~I.~Zakharov,
  %``Can Confinement Ensure Natural CP Invariance of Strong Interactions?,''
  Nucl.\ Phys.\ B {\bf 166} (1980) 493.
  %%CITATION = NUPHA,B166,493;%%  

\bibitem{Dine:1981rt}
  M.~Dine, W.~Fischler and M.~Srednicki,
  %``A Simple Solution to the Strong CP Problem with a Harmless Axion,''
  Phys.\ Lett.\ B {\bf 104} (1981) 199.
  %%CITATION = PHLTA,B104,199;%%

\bibitem{Zhitnitsky:1980tq}
  A.~R.~Zhitnitsky,
  %``On Possible Suppression of the Axion Hadron Interactions. (In Russian),''
  Sov.\ J.\ Nucl.\ Phys.\  {\bf 31} (1980) 260
   [Yad.\ Fiz.\  {\bf 31} (1980) 497].
  %%CITATION = SJNCA,31,260;%%

\bibitem{Kim:1998va}
  J.~E.~Kim,
  %``Constraints on very light axions from cavity experiments,''
  Phys.\ Rev.\ D {\bf 58} (1998) 055006
  [hep-ph/9802220].
  %%CITATION = HEP-PH/9802220;%%

\bibitem{Martin:1997}
  S.~P.~Martin,
  %``A Supersymmetry primer,''
  in *Kane, G.L. (ed.): Perspectives on supersymmetry II* 1-153
  [hep-ph/9709356].
  %%CITATION = HEP-PH/9709356;%%

\bibitem{Choi:2013}
  K.-Y.~Choi, J.~E.~Kim, L.~Roszkowski,
  %``Review of axino dark matter,''
  J.\ Korean Phys.\ Soc.\ 63 (2013) 1685-1695
  arXiv:1307.3330 [astro-ph.CO].
  %%CITATION = ASTRO-PH.CO/1307.3330;%%

\bibitem{Nilles:1981py}
  H.~P.~Nilles and S.~Raby,
  %``Supersymmetry and the strong CP problem,''
  Nucl.\ Phys.\ B {\bf 198} (1982) 102.
  %%CITATION = NUPHA,B198,102;%%

\bibitem{Tamvakis:1982mw}
  K.~Tamvakis and D.~Wyler,
  %``Broken Global Symmetries in Supersymmetric Theories,''
  Phys.\ Lett.\ B {\bf 112} (1982) 451.
  
\bibitem{Frere:1982sg}
  J.~M.~Frere and J.~M.~Gerard,
  %``Axions and Supersymmetry,''
  Lett.\ Nuovo Cim.\  {\bf 37} (1983) 135.
  %%CITATION = NCLTA,37,135;%%
  
\bibitem{Covi:1999ty}
  L.~Covi, J.~E.~Kim and L.~Roszkowski,
  %``Axinos as cold dark matter,''
  Phys.\ Rev.\ Lett.\  {\bf 82} (1999) 4180
  [hep-ph/9905212].
  %%CITATION = HEP-PH/9905212;%%

\bibitem{Covi:2001nw}
  L.~Covi, H.~B.~Kim, J.~E.~Kim and L.~Roszkowski,
  %``Axinos as dark matter,''
  JHEP {\bf 0105} (2001) 033
  [hep-ph/0101009].
  %%CITATION = HEP-PH/0101009;%%

\bibitem{Chun:1992zk}
  E.~J.~Chun, J.~E.~Kim and H.~P.~Nilles,
  %``Axino mass,''
  Phys.\ Lett.\ B {\bf 287} (1992) 123
  [hep-ph/9205229].
  %%CITATION = HEP-PH/9205229;%%

\bibitem{Chun:1995hc}
  E.~J.~Chun and A.~Lukas,
  %``Axino mass in supergravity models,''
  Phys.\ Lett.\ B {\bf 357} (1995) 43
  [hep-ph/9503233].
  %%CITATION = HEP-PH/9503233;%%

\bibitem{Kim:2012bb}
  J.~E.~Kim and M.~S.~Seo,
  %``Mixing of axino and goldstino, and axino mass,''
  Nucl.\ Phys.\ B {\bf 864} (2012) 296
  [arXiv:1204.5495 [hep-ph]].
  %%CITATION = ARXIV:1204.5495;%%

\bibitem{Covi:2009pq}
  L.~Covi and J.~E.~Kim,
  %``Axinos as Dark Matter Particles,''
  New J.\ Phys.\  {\bf 11} (2009) 105003
  [arXiv:0902.0769 [astro-ph.CO]].
  %%CITATION = ARXIV:0902.0769;%%

\bibitem{Brandenburg:2004du}
  A.~Brandenburg and F.~D.~Steffen,
  %``Axino dark matter from thermal production,''
  JCAP {\bf 0408} (2004) 008
  [hep-ph/0405158].
  %%CITATION = HEP-PH/0405158;%%

\bibitem{Chun:2011zd}
  E.~J.~Chun,
  %``Dark matter in the Kim-Nilles mechanism,''
  Phys.\ Rev.\ D {\bf 84} (2011) 043509
  [arXiv:1104.2219 [hep-ph]].
  %%CITATION = ARXIV:1104.2219;%%

  \bibitem{Giudice:2000}
  G.~F.~Giudice, E.~W.~Kolb, A.~Riotto,
  %``Largest temperature of the radiation era and its cosmological implications ,''
  Phys.\ Rev.\ D 64 (2001) 023508
  [hep-ph/0005123].
  %%CITATION = HEP-PH/0005123;%%

\bibitem{Fornengo:2002}
  N.~Fornengo, A.~Riotto, S.~Scopel,
  %``Supersymmetric Dark Matter and the Reheating Temperature of the Universe,''
  Phys.\ Rev.\ D 67 (2003) 023514
  [hep-ph/0208072].
  %%CITATION = HEP-PH/0208072;%%

\bibitem{Gelmini:2006Feb}
  G.~Gelmini, P.~Gondolo,
  %``Neutralino with the Right Cold Dark Matter Abundance in (Almost) Any Supersymmetric Model,''
  Phys.\ Rev.\ D 74 (2006) 023510
  [hep-ph/0602230].
  %%CITATION = HEP-PH/0602230;%%

\bibitem{Gelmini:2006May}
  G.~Gelmini, G.~Gondolo, A.~Soldatenko, C.~E.~Yaguna,
  %``The effect of a late decaying scalar on the neutralino relic density,''
  Phys.\ Rev.\ D 74 (2006) 083514
  [hep-ph/0605016].
  %%CITATION = HEP-PH/0605016;%%
  
\bibitem{Gelmini:2006Oct}
  G.~B.~Gelmini, G.~Gondolo, A.~Soldatenko, C.~E.~Yaguna,
  %``Direct detection of neutralino dark mattter in non-standard cosmologies,''
  Phys. Rev.\ D 76 (2007) 015010
  [hep-ph/0610379].
  %%CITATION = HEP-PH/0610379;%%

\bibitem{Strumia:2010aa}
  A.~Strumia,
  %``Thermal production of axino Dark Matter,''
  JHEP {\bf 1006} (2010) 036
  [arXiv:1003.5847 [hep-ph]].
  %%CITATION = ARXIV:1003.5847;%%
  
\bibitem{Roszkowski:2014lga}
  L.~Roszkowski, S.~Trojanowski and K.~Turzynski,
  %``Neutralino and gravitino dark matter with low reheating temperature,''
  JHEP {\bf 1411} (2014) 146
  [arXiv:1406.0012 [hep-ph]].
  %%CITATION = ARXIV:1406.0012;%%


\bibitem{Co:2015pka}
  R.~T.~Co, F.~D'Eramo, L.~J.~Hall and D.~Pappadopulo,
  %``Freeze-In Dark Matter with Displaced Signatures at Colliders,''
  arXiv:1506.07532 [hep-ph].
  %%CITATION = ARXIV:1506.07532;%%


\bibitem{Choi:2011yf}
  K.-Y.~Choi, L.~Covi, J.~E.~Kim and L.~Roszkowski,
  %``Axino Cold Dark Matter Revisited,''
  JHEP {\bf 1204} (2012) 106
  [arXiv:1108.2282 [hep-ph]].
  %%CITATION = ARXIV:1108.2282;%%

\bibitem{Monteux:2015qqa}
  A.~Monteux and C.~S.~Shin,
  %``Thermal Goldstino Production with Low Reheating Temperatures,''
  arXiv:1505.03149 [hep-ph].

\bibitem{Kawasaki:1999na}
  M.~Kawasaki, K.~Kohri and N.~Sugiyama,
  %``Cosmological constraints on late time entropy production,''
  Phys.\ Rev.\ Lett.\  {\bf 82} (1999) 4168
  [astro-ph/9811437].
  %%CITATION = ASTRO-PH/9811437;%%

\bibitem{Kawasaki:2000en}
  M.~Kawasaki, K.~Kohri and N.~Sugiyama,
  %``MeV scale reheating temperature and thermalization of neutrino background,''
  Phys.\ Rev.\ D {\bf 62} (2000) 023506
  [astro-ph/0002127].
  %%CITATION = ASTRO-PH/0002127;%%

\bibitem{Roszkowski:2004jd}
  L.~Roszkowski, R.~Ruiz de Austri and K.-Y.~Choi,
  %``Gravitino dark matter in the CMSSM and implications for leptogenesis and the LHC,''
  JHEP {\bf 0508} (2005) 080
  [hep-ph/0408227].
  
\bibitem{Cerdeno:2005eu}
  D.~G.~Cerdeno, K.-Y.~Choi, K.~Jedamzik, L.~Roszkowski and R.~Ruiz de Austri,
  %``Gravitino dark matter in the CMSSM with improved constraints from BBN,''
  JCAP {\bf 0606} (2006) 005
  [hep-ph/0509275].
  %%CITATION = HEP-PH/0509275;%%
  %126 citations counted in INSPIRE as of 27 May 2014

\bibitem{Steffen:2006}
  F.~D.~Steffen,
  %``Gravitino Dark Matter and Cosmological Constraints ,''
  JCAP 0609 (2006) 001
  [hep-ph/0605306].
  %%CITATION = HEP-PH/0605306;%%

\bibitem{Kanzaki:2006}
  T.~Kanzaki, M.~Kawasaki, K.~Kohri, T.~Moroi, 
  %``Cosmological Constraints on Gravitino LSP Scenario with Sneutrino NLSP,''
  Phys.\ Rev.\ D 75 (2007) 025011
  [hep-ph/0609246].
  %%CITATION = HEP-PH/0609246;%%

%\cite{Pospelov:2006sc}
\bibitem{Pospelov:2006sc}
  M.~Pospelov,
  %``Particle physics catalysis of thermal Big Bang Nucleosynthesis,''
  Phys.\ Rev.\ Lett.\  {\bf 98} (2007) 231301
  [hep-ph/0605215].
  %%CITATION = HEP-PH/0605215;%%

%\cite{Jedamzik:2007qk}
\bibitem{Jedamzik:2007qk}
  K.~Jedamzik,
  %``Bounds on long-lived charged massive particles from Big Bang nucleosynthesis,''
  JCAP {\bf 0803} (2008) 008
  [arXiv:0710.5153 [hep-ph]].
  %%CITATION = ARXIV:0710.5153;%%

%\cite{Kawasaki:2007xb}
\bibitem{Kawasaki:2007xb}
  M.~Kawasaki, K.~Kohri and T.~Moroi,
  %``Big-Bang Nucleosynthesis with Long-Lived Charged Slepton,''
  Phys.\ Lett.\ B {\bf 649}, 436 (2007)
  [hep-ph/0703122].
  %%CITATION = HEP-PH/0703122;%%


\bibitem{Kawasaki:2008qe}
  M.~Kawasaki, K.~Kohri, T.~Moroi and A.~Yotsuyanagi,
  %``Big-Bang Nucleosynthesis and Gravitino,''
  Phys.\ Rev.\ D {\bf 78}, 065011 (2008)
  [arXiv:0804.3745 [hep-ph]].
  %%CITATION = ARXIV:0804.3745;%%
  
 \bibitem{Covi:2002vw}
  L.~Covi, L.~Roszkowski and M.~Small,
  %``Effects of squark processes on the axino CDM abundance,''
  JHEP {\bf 0207} (2002) 023
  [hep-ph/0206119].
  %%CITATION = HEP-PH/0206119;%%

\bibitem{Choi:2012zna}
  K.~Choi, K.-Y.~Choi and C.~S.~Shin,
  %``Dark radiation and small-scale structure problems with decaying particles,''
  Phys.\ Rev.\ D {\bf 86} (2012) 083529
  [arXiv:1208.2496 [hep-ph]].
  %%CITATION = ARXIV:1208.2496;%%
  %31 citations counted in INSPIRE as of 28 Sep 2015

\bibitem{Hisano:2000dz}
  J.~Hisano, K.~Kohri and M.~M.~Nojiri,
  %``Neutralino warm dark matter,''
  Phys.\ Lett.\ B {\bf 505} (2001) 169
  [hep-ph/0011216].
  %%CITATION = HEP-PH/0011216;%%
  %56 citations counted in INSPIRE as of 28 Sep 2015

\bibitem{Jedamzik:2005sx}
  K.~Jedamzik, M.~Lemoine and G.~Moultaka,
  %``Gravitino, axino, Kaluza-Klein graviton warm and mixed dark matter and reionisation,''
  JCAP {\bf 0607} (2006) 010
  [astro-ph/0508141].
  %%CITATION = ASTRO-PH/0508141;%%
  %66 citations counted in INSPIRE as of 28 Sep 2015

\bibitem{Boyarsky:2008xj}
  A.~Boyarsky, J.~Lesgourgues, O.~Ruchayskiy and M.~Viel,
  %``Lyman-alpha constraints on warm and on warm-plus-cold dark matter models,''
  JCAP {\bf 0905} (2009) 012
  [arXiv:0812.0010 [astro-ph]].
  %%CITATION = ARXIV:0812.0010;%%

 \bibitem{Covi:2009}
  L.~Covi, J.~Hasenkamp, S.~Pokorski, J.~Roberts,
  %``Gravitino Dark Matter and general neutralino NLSP,''
  JHEP 0911 (2009) 003
  arXiv:0908.3399 [hep-ph].
  %%CITATION = HEP-PH/0908.3399;%%

\bibitem{Aad:2014yka}
  G.~Aad {\it et al.} [ATLAS Collaboration],
  %``Search for the direct production of charginos, neutralinos and staus in final states with at least two hadronically decaying taus and missing transverse momentum in $pp$ collisions at $\sqrt{s}$ = 8 TeV with the ATLAS detector,''
  JHEP {\bf 1410} (2014) 96
  [arXiv:1407.0350 [hep-ex]].
  %%CITATION = ARXIV:1407.0350;%%

\bibitem{Alwall:2014hca}
  J.~Alwall {\it et al.},
  %``The automated computation of tree-level and next-to-leading order differential cross sections, and their matching to parton shower simulations,''
  JHEP {\bf 1407} (2014) 079
  [arXiv:1405.0301 [hep-ph]].
  %%CITATION = ARXIV:1405.0301;%%

\bibitem{Mukaida:2015ria}
  K.~Mukaida and M.~Yamada,
  %``Thermalization Process after Inflation and Effective Potential of Scalar Field,''
  arXiv:1506.07661 [hep-ph].
  %%CITATION = ARXIV:1506.07661;%%

\bibitem{Affleck:1984fy}
  I.~Affleck and M.~Dine,
  %``A New Mechanism for Baryogenesis,''
  Nucl.\ Phys.\ B {\bf 249} (1985) 361.
  %%CITATION = NUPHA,B249,361;%%

\bibitem{Allahverdi:2012ju}
  R.~Allahverdi and A.~Mazumdar,
  %``A mini review on Affleck-Dine baryogenesis,''
  New J.\ Phys.\  {\bf 14} (2012) 125013.
  %%CITATION = NJOPF,14,125013;%%

\bibitem{Bae:2011jb}
  K.~J.~Bae, K.~Choi and S.~H.~Im,
  %``Effective Interactions of Axion Supermultiplet and Thermal Production of Axino Dark Matter,''
  JHEP {\bf 1108} (2011) 065
  [arXiv:1106.2452 [hep-ph]].
  %%CITATION = ARXIV:1106.2452;%%

\bibitem{Kim:1983dt}
  J.~E.~Kim and H.~P.~Nilles,
  %``The mu Problem and the Strong CP Problem,''
  Phys.\ Lett.\ B {\bf 138} (1984) 150.
  %%CITATION = PHLTA,B138,150;%%
 
 \bibitem{Choi:1999xm}
  K.~Choi, K.~Hwang, H.~B.~Kim and T.~Lee,
  %``Cosmological gravitino production in gauge mediated supersymmetry breaking models,''
  Phys.\ Lett.\ B {\bf 467} (1999) 211
  [hep-ph/9902291].
  %%CITATION = HEP-PH/9902291;%%
  
\bibitem{PDG:2014}
  J.~Beringer \textit{et al.} (Particle Data Group), 
  Phys.\ Rev.\ D {86}, 010001 (2012) and 2013 partial update for the 2014 edition

\bibitem{Fowlie:2013}
  A.~Fowlie, K.~Kowalska, L.~Roszkowski, E.~M.~Sessolo, Y.-L.~S.~Tsai,
  %``Dark matter and collider signatures of the MSSM,''
  Phys.\ Rev.\ D 88 (2013) 055012
  arXiv:1306.1567 [hep-ph].
  %%CITATION = HEP-PH/1306.1567;%%
  
\bibitem{Feroz:2009}
  F.~Feroz, M.~Hobson, M.~Bridges,
  %``MultiNest: an efficient and robust Bayesian inference tool for cosmology and particle physics,''
  Mon.\ Not.\ Roy.\ Astron.\ Soc.\ 398 (2009) 1601-1614
  arXiv:0809.3437 [astro-ph].
  %%CITATION = HEP-PH/0809.3437;%%
  
\bibitem{Allanach:2002}
  B.~Allanach, 
  %``SOFTSUSY: a program for calculating supersymmetric spectra ,''
  Comput.\ Phys.\ Commun.\ 143 (2002) 305-331
  [hep-ph/0104145].
  %%CITATION = HEP-PH/0104145;%%
  
\bibitem{Arbey:2009}
  A.~Arbey, F.~Mahmoudi, 
  %``SuperIso Relic: A program for calculating relic density and flavor physics observables in Supersymmetry,''
  Comput.\ Phys.\ Commun.\ 181 (2010) 1277-1292
  arXiv:0906.0369 [hep-ph].
  %%CITATION = HEP-PH/0906.0369;%%
   
\bibitem{CMS:2013}
  \textbf{CMS} Collaboration, 
  %``Measurements of the properties of the new boson with a mass near 125 GeV,''
  Tech.\ Rep.\ CMS-PAS-HIG-13-005, CERN, Geneva, 2013.

\bibitem{Planck:2013}
  \textbf{Planck} Collaboration, P.~Ade \textit{et al.},
  %``Planck 2013 results. XVI. Cosmological parameters,''
  arXiv:1303.5076 [astro-ph.CO].
  %%CITATION = ASTR-PH.CO/1303.5076;%%

\bibitem{SLAC}
    http://www.slac.stanford.edu/xorg/hfag/rare/2012/radll/index.ht

\bibitem{Belle:2012}
  \textbf{Belle} Collaboration, I.~Adachi \textit{et al.},
  %``Evidence for B- -> tau- nu_tau-bar with a Hadronic Tagging Method Using the Full Data Sample of Belle,''
  Phys.\ Rev.\ Lett.\ 110 (2013) 131801
  arXiv:1208.4678 [hep-ex].
  %%CITATION = HEP-EX/1208.4678;%%
 
\bibitem{LHCb:2013}
  \textbf{LHCb} Collaboration, R.~Aaij \textit{et al.},
  %``Measurement of the B0sâÎ¼+Î¼â branching fraction and search for B0âÎ¼+Î¼â decays at the LHCb experiment,''
  Phys.\ Rev.\ Lett.\ 111 (2013) 101805
  arXiv:1307.5024 [hep-ex].
  %%CITATION = HEP-EX/1307.5024;%%

\bibitem{CMS:2013Jul}
  \textbf{CMS} Collaboration, S.~Chatrchyan \textit{et al.},
  %``Measurement of the B(s) to mu+ mu- branching fraction and search for B0 to mu+ mu- with the CMS Experiment,''
  Phys.\ Rev.\ Lett.\ 111 (2013) 101804
  arXiv:1307.5025 [hep-ex].
  %%CITATION = HEP-EX/1307.5025;%%

\bibitem{LUX:2013}
  \textbf{LUX} Collaboration, D.~S.~Akerib \textit{et al.},
  %``First results from the LUX dark matter experiment at the Sanford Underground Research Facility,''
  Phys.\ Rev.\ Lett.\ 112 (2014) 091303
  arXiv:1310.8214 [astro-ph.CO].
  %%CITATION = ASTRO-PH.CO/1310.8214;%%  

\bibitem{Fowlie:2012}
  A.~Fowlie, M.~Kazana, K.~Kowalska, S.~Munir, L.~Roszkowski, E.~M.~Sessolo, S.~Trojanowski, Y.-L.~S.~Tsai,
  %``Constrained MSSM favoring new territories: The impact of new LHC limits and a 125 GeV Higgs boson,''
  Phys.\ Rev.\ D 86 (2012) 075010
  arXiv:1206.0264 [hep-ph].
  %%CITATION = HEP-PH/1206.0264;%%

\bibitem{Dev:2013yza}
  P.~S.~Bhupal Dev, A.~Mazumdar and S.~Qutub,
  %``Constraining Non-thermal and Thermal properties of Dark Matter,''
  Front.\ Phys.\  {\bf 2} (2014) 26
  [arXiv:1311.5297 [hep-ph]].
  %%CITATION = ARXIV:1311.5297;%%
  %13 citations counted in INSPIRE as of 21 Jul 2015

\bibitem{Belanger:2013}
  B.~Belanger, F.~Boudjema, A.~Pukhov, A.~Semenov,
  %``micrOMEGAs3.1 : a program for calculating dark matter observables,''
  Comput.\ Phys.\ Commun.\ 185 (2014) 960-985
  arXiv:1305.0237 [hep-ph].
  %%CITATION = HEP-PH/1305.0237;%%

                    
\end{thebibliography}
\end{document}